\documentclass[twocolumn,secnumarabic,amssymb,superscriptaddress,nobibnotes, aps, prd]{revtex4-2}

\usepackage{comment}
\usepackage{graphicx}
\usepackage[charter,greekuppercase=italicized]{mathdesign}
\usepackage{amsmath, amsfonts, amssymb}
\usepackage{hyperref}
\usepackage{xcolor}
\hypersetup{
    colorlinks = true,
    linkcolor = blue,
    anchorcolor = blue,
    citecolor = blue,
    filecolor = blue,
    urlcolor = blue
    }   
\usepackage{multirow}
\usepackage{hhline}
\usepackage{xcolor}

\begin{document}

\title{Hydrostatic and chemical pressure driven crossover from commensurate to the incommensurate state of the Weyl semimetal Mn$_{3+x}$Sn$_{1-x}$}%

\author{K. Bhattacharya}
\address{Department of Physics, Shiv Nadar Institution of Eminence, Gautam Buddha Nagar, UP 201314, India}

\author{A. K. Bharatwaj}
\address{Department of Physics, Shiv Nadar Institution of Eminence, Gautam Buddha Nagar, UP 201314, India}

\author{C. Singh}
\address{School of Physical Sciences, National Institute of Science Education and Research, HBNI, Jatni 752050, India}
\address{Physikalisches Institut, Karlsruhe Institute of Technology, 76131 Karlsruhe, Germany}

\author{R. Gupta}
\address{Laboratory for Muon Spin Spectroscopy, Paul Scherrer Institut, 5232 Villigen PSI, Switzerland}
\address{Department of Physics, Indian Institution of Technology Ropar, Punjab-140001, India}

\author{ R. Khasanov}
\address{Laboratory for Muon Spin Spectroscopy, Paul Scherrer Institut, 5232 Villigen PSI, Switzerland}

\author{S. Kanungo}
\address{School of Physical Sciences, Indian Institute of Technology Goa, Goa-403401, India}

\author{A. K. Nayak}
\address{School of Physical Sciences, National Institute of Science Education and Research, HBNI, Jatni 752050, India}

\author{M. Majumder}
\email[Corresponding author: ]{mayukh.majumder@snu.edu.in}
\address{Department of Physics, Shiv Nadar Institution of Eminence, Gautam Buddha Nagar, UP 201314, India}

\date{\today}
\begin{abstract}
The observation of large intrinsic anomalous Hall conductivity (AHC) in the non-collinear antiferromagnetic (AFM) phase of the Weyl semimetal Mn$_3$Sn generates enormous interest in uncovering the entanglement between the real space magnetic ordering and the momentum space band structure. Previous studies show that changes in the magnetic structure induced by the application of hydrostatic and chemical pressure can significantly affect the  AHC of Mn$_{3+x}$Sn$_{1-x}$ system.  Here, we employ the muon spin relaxation/rotation ($\mu^+$SR) technique to systematically investigate the evolution of different magnetic states in the  Mn$_{3+x}$Sn$_{1-x}$ as a function of hydrostatic and chemical pressure. We find two muon sites experimentally, which is also supported by our \textit{ab initio} calculations. Our $\mu^+$SR experiments affirm that the $x = 0.05$ compound exhibits a commensurate magnetic state throughout the magnetically ordered phase below the Neel temperature $T_N \approx 420$~K in ambient pressure.  In contrast, we observe an incommensurate magnetic state below $T_{IC} \sim 175$~K when a hydrostatic pressure of 1.5~GPa is applied. A similar transition from the commensurate to incommensurate state is also found with chemical pressure for  $x = 0.04$ and $x = 0.03$, using $\mu^+$SR and elastic neutron scattering experiments.  Using band structure calculations, we have shown the emergence of Fermi nesting in Mn$_3$Sn and the subsequent development of incommensurate magnetic ordering under hydrostatic/chemical pressure.

\end{abstract}

\maketitle

\section{Introduction}
Weyl semimetals (WSM) \cite{PhysRevX.5.031013, doi:10.1126/science.aaa9297, Yang2015} are topological insulators \cite{RevModPhys.83.1057, RevModPhys.82.3045} with gapless linear dispersion, even if their surface states are not shielded against backscattering because of time-reversal symmetry ($\mathcal{T}$) and/or inversion symmetry ($\mathcal{P}$) breaking. This mechanism contributes to the emergence of several transport anomalies, which are fascinating since we may utilize them for ambient temperature applications. One such well-known Weyl semimetal is Mn$_3$Sn \cite{PhysRevB.83.205101, PhysRevLett.107.127205} that exhibits several interesting transport phenomena, such as large anomalous Hall effect (AHE) at room temperature \cite{Nakatsuji2015}, topological Hall effect (THE) \cite{PhysRevB.99.094430},  spin Hall effect (SHE) \cite{Kimata2019}, and large magneto-optical Kerr effect \cite{Higo2018} (MOKE), etc. Mn$_3$Sn crystallizes in a hexagonal structure (similar to Mn$_3$X [X = Ga, Ge] with the space group $P6_3/mmc$) with a kagome configuration of Mn-atoms in the \textit{a-b} plane where the magnetic moment of each Mn atom is about 3 $\mu_B$ \cite{P_J_Brown_1990, Tomiyoshi_Yamaguchi_1982,PhysRevB.101.144422, wang2023flat}. The neutron scattering experiments have shown that it has a non-collinear 120$^{\circ}$ inverse triangular antiferromagnetic (iT-AFM) order with a modulation wavevector $\textbf{k}= 0$ below $T_N \approx 420$~K driven by geometrical frustration of nearest-neighbor antiferromagnetic and Dzyaloskinskii-Moriya (DM) interaction mediated by spin-orbit coupling (SOC) \cite{Park2018, Tomiyoshi_Yamaguchi_1982}. 

Theoretically it has been established that non-collinear 120$^\circ$ iT-AFM spin structure of Mn$_3$Sn is responsible for a large anomalous Hall conductivity (AHC) induced by Berry curvature mechanism \cite{PhysRevB.95.075128,Kübler_2014}. Based on extensive electronic structure calculations and low energy \textit{\textbf{k.p}} effective theory with various orientations of $120^\circ$ noncollinear iT-AFM spin structures driven vector chirality, Pradhan \textit{et al.} have shown that transition from nodal-ring band-structure to Weyl nodes can lead to switching of AHC from zero to a finite value \cite{Pradhan2023}. Plenty of practical applications like magnetic memory devices, spintronics, quantum computing, thermoelectric devices, and tensor technologies, etc. may be made possible by this kind of AHC ($\sigma_{xy}^A$) tuning, which involves modulation of its magnetic state in different ways. Tuning of AHC through external stimuli like physical pressure and doping has been observed in several compounds.
 
For example, a Van der Waals ferromagnet  Fe$_3$GeTe$_2$ and the antiferromagnet half-Heusler compound GdPtBi exhibit decreasing $\sigma ^A _{xy}$ with increasing hydrostatic pressure \cite{PhysRevB.100.014407, PhysRevB.103.085116}. On the other hand, one can tune $\sigma ^A _{xy}$ by altering the stoichiometric ratio by doping the system intrinsically \cite{PhysRevB.107.184413, PhysRevLett.125.086602, 10.1063/5.0095950, Liu2018}. Similarly, Guguchia \textit{et al.} have investigated Co$_3$Sn$_2$S$_2$ using muon spin relaxation/ rotation ($\mu ^+$SR) experiment and showed that pressure change may be utilized to control the AHC through adjusting the competition between the FM and AFM states \cite{Guguchia2020}.

Experimentally the tuning of AHC has also been observed in Mn$_3$Sn using bulk measurements. It has been observed that another magnetic phase transition in Mn$_3$Sn can occur between 175 - 290~K temperature range depending on their synthesis procedure and Mn-concentration, as described in plenty of literature \cite{KPZZ1975,PhysRevB.101.144422, cao2023optical, doi:10.7566/JPSCP.30.011177, Nayak2020} which may lead to the change in AHC. In general, the kagome structure in metals gives an ideal playground to correlate itinerant-electron, magnetism, band topology, Weyl nodes, flat bands, and Van Hove Singularities which creates a perfect Fermi nesting condition that may trigger instabilities for spin and charge associated with some characteristic modulated wave vectors \cite{Hu2022, PhysRevB.97.094412, Kang2020}. This kind of mechanism of spin density wave (SDW) and charge density wave (CDW) is commonly studied in the field of superconductors \cite{PhysRevB.103.104501, PhysRevB.81.174538, Tam2022}. These instabilities may have caused the system to become relaxed by introducing a new type of magnetic order. Theoretical studies predicted that at low temperatures (below 175 - 290~K), the nature of the magnetic ordering in Mn$_3$Sn is a canted spin spiral helical magnetic structure (SDW-type) different than high-temperature non-collinear iT-AFM \cite{PhysRevB.99.094430, Park2018}. Neutron diffraction studies also corroborate these predictions \cite{Park2018, PhysRevResearch.6.L032016}.

Similar transition in this temperature range has been observed under hydrostatic pressure conditions using bulk measurements \cite{Nayak2020}. Interestingly, AHC also disappears suddenly below this temperature \cite {Nayak2020}.  However, until now no microscopic tool has been used to systematically study the type of magnetic ordering introduced by the means of physical hydrostatic pressure and chemical doping in Mn$_3$Sn. As a result, it is crucial to conduct an experimental investigation to explore the nature of magnetic states of Mn$_3$Sn with the physical pressure and changing the stoichiometric ratio, which would lead to the elucidation of underlying mechanisms for the presence or absence of AHC. In this work, we employed $\mu ^+$SR and neutron diffraction technique to understand the evolution of the magnetic ground state of Mn$_{3+x}$Sn$_{1-x}$ for different values of $x$ and under hydrostatic pressure. We have also carried out \textit{first-principles} calculations to investigate the mechanism behind this evolution of the magnetic ground state under hydrostatic pressure in a systematic manner.


\begin{figure}
\includegraphics[scale=0.22]{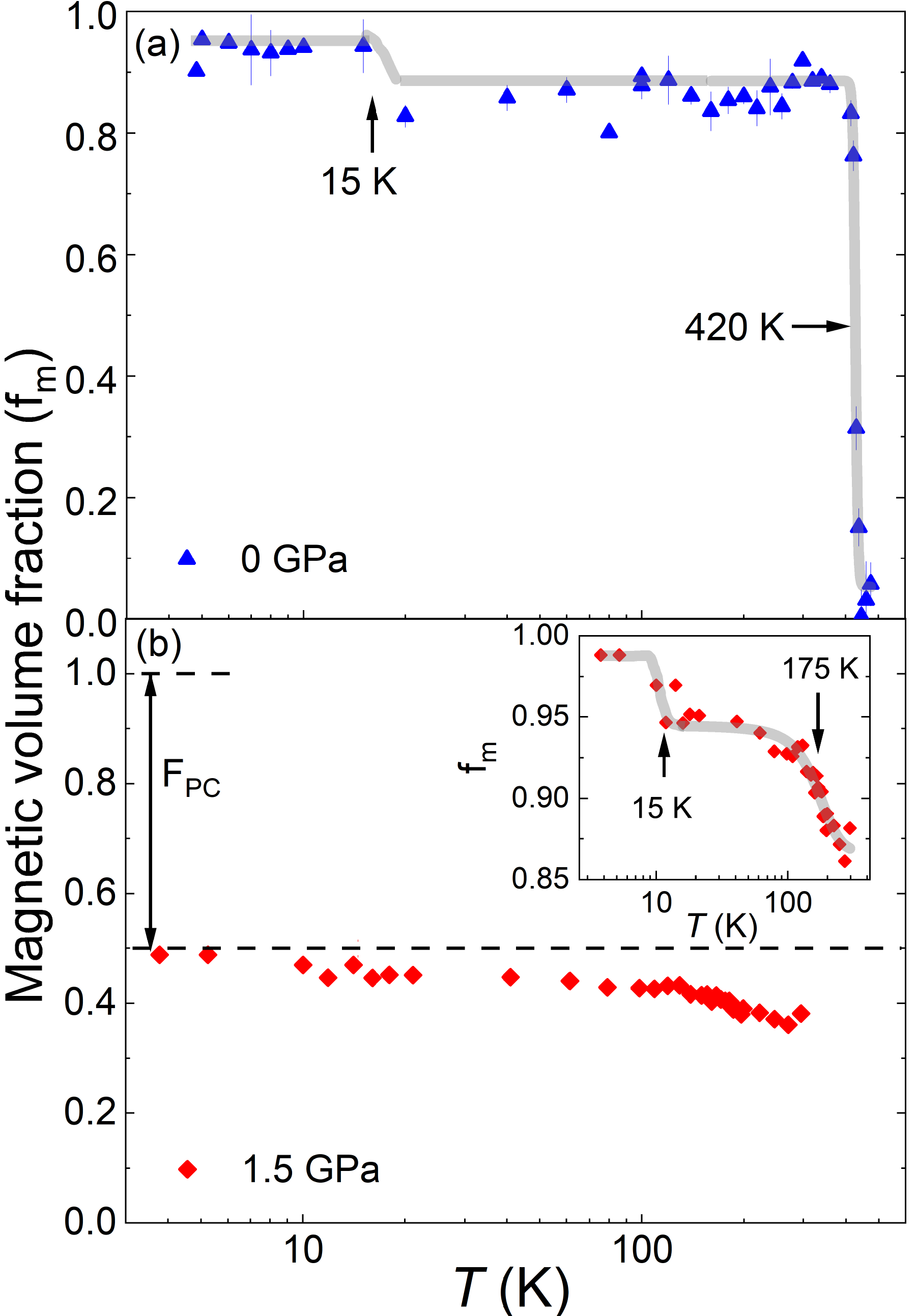}
\caption{\label{wTF_ZP_GPD}Magnetic volume fraction $(f_m)$ obtained from the wTF measurements for Mn$_{3.05}$Sn$_{0.95}$ at \textbf{(a)} ambient and \textbf{(b)} 1.5~GPa pressure with pressure cell contribution. Inset of \textbf{(b)} shows a zoomed-in version of $f_m$ versus temperature at 1.5~GPa pressure after subtracting pressure cell contribution.}
\end{figure}

\section{Experimental Details}
Three distinct polycrystalline ingots of Mn$_{3+x}$Sn$_{1-x}$ (with $x=0.05, 0.04,$ and $0.03$) were used for $\mu^+$SR measurements to investigate the nature of magnetic states. The same batch of samples prepared for the work in Ref.~\cite{Nayak2020} have been used for this study. These stoichiometric ratios have been determined by the use of Energy-dispersive X-ray analysis (EDAX) measurement though utilizing EDAX to verify such minor modifications is challenging. Nevertheless, after preparation, multiple batches of the same Mn-concentration compounds were discovered to have intact initial compositions and expected magnetic characteristics. The $\mu$SR asymmetry is defined as the temporal evolution of the normalized difference between the angle-resolved positron numbers. These are sensed by detectors via the implanted interstitial fully polarized muon ($\mu^+$) spins decay, providing significant information about the magnetism in the system at the microscopic level. We performed the $\mu^+$SR experiment without an externally applied magnetic field, known as zero-field (ZF) measurement. The precession frequency of the muon spins is proportional to the local intrinsic magnetic field at the muon sites in the interstitial position of the lattice. Alongside ZF, we have also conducted weak transverse field (wTF) measurements at 50G for various temperatures, where the initial asymmetry directly related to the non-magnetic volume fraction within the compound. These experiments were carried out at the Swiss muon source (S$\mu$S) in Paul Scherrer Institute (PSI), Villigen, Switzerland using the GPS spectrometer at $\pi$M3.2 beamline for measurements  at ambient pressure in the temperature range of 475~K to 5~K (wTF measurement with 475 - 5~K and 300 - 50~K for ZF measurements). For under-pressure measurements in the temperature range of 275~K to 50~K (295 - 4~K for wTF measurements and 275 - 50~K for ZF measurements), the GPD spectrometer was used, placed at the $\mu$E1 area, where an intense high-energy beam of muons was implanted into the sample through the pressure cell. The pressure cell was made of a double-wall CuBe/MP35N piston cylinder, and pressure was then transmitted to the powdered sample using 7373 Daphne oil as a pressure-transmitting medium \cite{doi:10.1080/08957959.2016.1173690, 10.1063/5.0119840, doi:10.1080/08957959.2017.1373773}. The pressure was determined by tracking the superconducting transition temperature of a tiny Indium (In) piece. All the results were obtained using the program MUSRFIT \cite{SUTER201269}.

The neutron powder diffraction (NPD) measurements of Mn$_{3.03}$Sn$_{0.97}$ at 300 - 50~K temperature range were carried out at the Dhruva Research Reactor of Bhabha Atomic Research Centre, Mumbai, India, using the neutron powder diffractometer (PD-1) \cite{Paranjpe1989, doi:10.1080/10448630208218701}. The NPD patterns were recorded across a 2$\theta$ angular range of $3 - 80^{\circ}$ with a step size of 0.05$^{\circ}$. The Rietveld analysis of experimentally recorded neutron diffraction patterns were carried out using the FULLPROF software \cite{Rodriguez1990}.

\section{RESULTS}
\subsection{Mn$_{3.05}$Sn$_{0.95}$}
\subsubsection{$\mu^+$SR-wTF measurements at ambient $\&$ 1.5~GPa pressure}
The wTF asymmetries as a function of time for the different temperatures at ambient and 1.5 GPa pressure for Mn$_{3.05}$Sn$_{0.95}$ have been  fitted with the following function

\begin{equation}
\begin{aligned}
\label{wtf}
	A_{wTF}(t) = A_0 f_{nm} cos(\omega t + \phi)	
\end{aligned}
\end{equation}

where $A_0, f_{nm}, \omega$ and $\phi$ are the initial asymmetry of the paramagnetic state, non-magnetic volume fraction, oscillation frequency, and initial offset phase respectively.  The magnetic volume fraction ($f_m=1-f_{nm}$) at ambient pressure as a function of temperature has been depicted in Fig.~\ref{wTF_ZP_GPD}(a). The observed magnetic volume fraction substantially increases at $T_N \sim$ 420~K by reducing the temperature, which confirms the Neel transition from the paramagnetic to iT-AFM state at ambient pressure. A reduced magnetic volume fraction between 400~K - 20~K, compared to the state below 15~K, indicates the presence of small amount of impurity. Whereas under the application of 1.5 GPa hydrostatic pressure, another transition arises in the region 175 - 200~K as shown in  Fig.\ref{wTF_ZP_GPD}(b). Note that, $f_m$ of the compound at under pressure measurement has pressure cell contribution as well. For clarity a zoomed-in figure without pressure cell contribution has been shown in the inset of Fig.\ref{wTF_ZP_GPD}(b), interestingly bulk measurements show that the AHC vanishes below this temperature region \cite{Nayak2020}. This is also consistent with the fact that the disappearance of AHE is related to the emergence of the new phase, as mentioned in several literature. The system also undergoes another transition below 15~K at ambient pressure and it is persistent under pressure as well. This transition was also observed in magnetization measurements \cite{kurosawa2022chiralanomalydriven, Nayak2020}, which was explained in terms out of plane canting of Mn moments that can give a finite topological Hall signal due to non-zero scalar spin chirality. Various groups in low-temperature regions have reported this type of transition \cite{PhysRevB.101.094404, PhysRevB.73.205105, PhysRevB.99.094430, singh2022higher}, which requires further studies to map out its origin. However, to enlighten the magnetic states corresponding to the vanishing of AHC below the high-temperature transition region, in the next sections, we analyze $\mu^+$SR data to understand better how the new magnetic state under pressure differs from the ambient pressure situation.

\subsubsection{$\mu^+$SR-ZF measurements at ambient pressure}
The observed spatial powder averaging $\mu$SR asymmetry at ambient pressure for Mn$_{3.05}$Sn$_{0.95}$ were fitted using the equation with a combination of two \textit{2/3-oscillatory} Cosine functions corresponding to two inequivalent muon sites (say, Site-I and Site-II, as our electrostatic potential calculations through DFT for possible muon sites also confirmed the existence of the two sites.) and two non-oscillatory \textit{1/3-tail} relaxing signals induced by field component fluctuations parallel to the original muon spin polarization,
  
\begin{equation}
\begin{aligned}
\label{3p05_ambient_pressure}
	A_{ZF}(t)=A_0 & \biggr[ \frac{2}{3} \sum_{i=1}^{2} f_i \cos(\omega_it+\phi)\exp^{-\lambda_{T_i}t} \\
				& +	\frac{1}{3} \sum_{i=1}^{2} \exp^{-\lambda_{L_i}t} \biggr]
\end{aligned}
\end{equation}

where $A_0$, $f_i$, $\omega_i(=2 \pi\nu_{\mu})$, $\phi$, $\lambda_{T_i}$, and $\lambda_{L_i}$ are the initial $\mu$SR asymmetry, the fractions of the sites, the Larmor frequencies corresponding to the sites, initial offset phases, transverse depolarization rates, and longitudinal depolarization rates respectively. The muon's Larmor frequency ($\nu_{\mu}$) is related to the average local field $B_{int}$  at the muon site with the following expression $2\pi\nu_{\mu} = \gamma_\mu B_{int}$  (with, $\gamma_\mu/2\pi$ = 135.5~MHz/T is the gyromagnetic ration of the $\mu^+$). 
    
\begin{figure}
\includegraphics[scale=0.4]{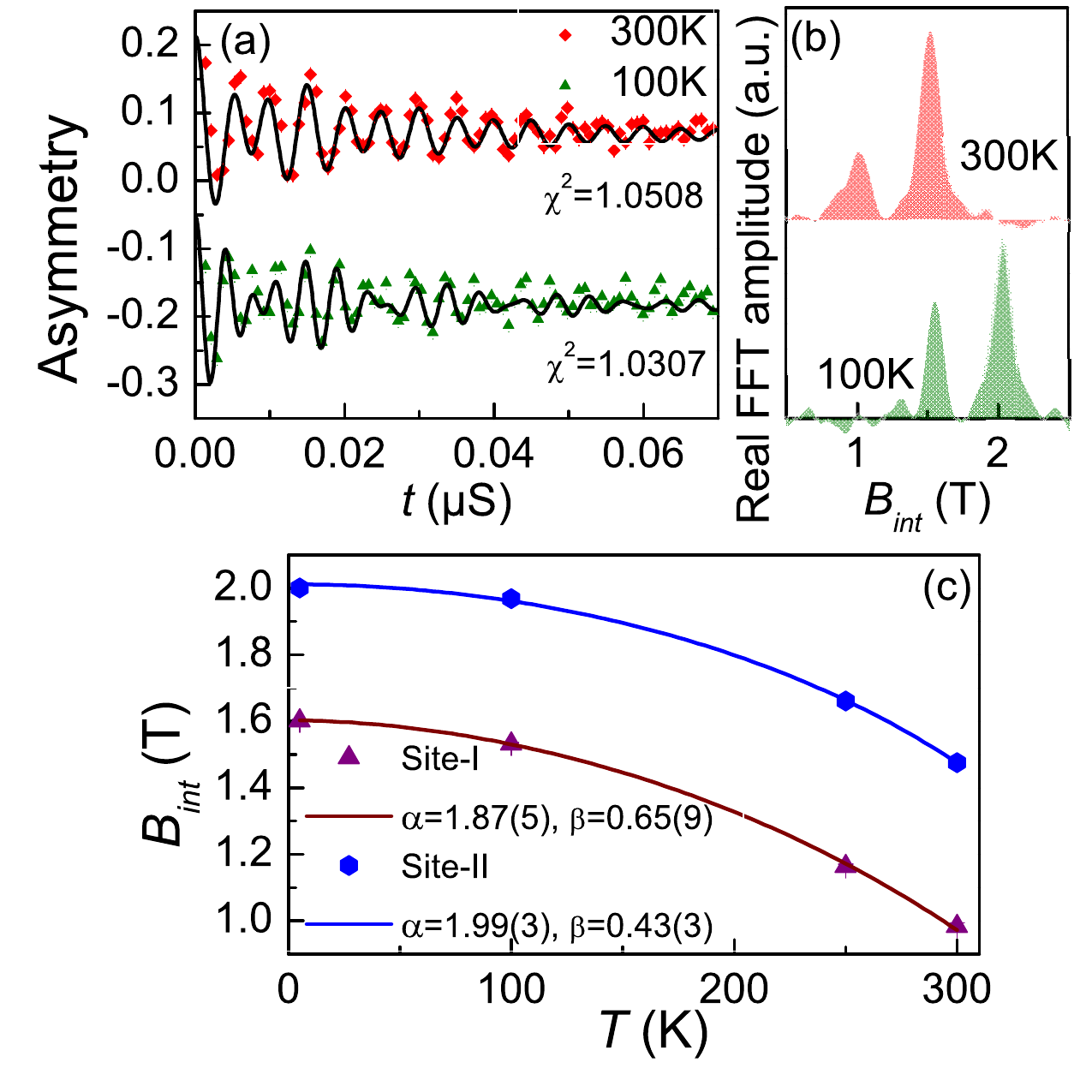}
\caption{\label{All_Final_3p05_ambient}Mn$_{3.05}$Sn$_{0.95}$ at ambient pressure: \textbf{(a)} $\mu$SR asymmetry at 300~K and 100~K (later spectrum is shifted downward by 0.25 for the clarity of display). \textbf{(b)} Fourier transformation amplitude of $\mu$SR asymmetry for 300~K and 100~K. \textbf{(c)} Internal magnetic field as a function of temperature for the two sites.}  
\end{figure}    
    
The $\mu$SR asymmetry for two temperatures 300~K and 100~K as a function of time is depicted in Fig. \ref{All_Final_3p05_ambient}(a). The two oscillatory components of the Eq.\ref{3p05_ambient_pressure} have high enough frequencies to squeeze the spectra into the early time domain ($<0.07$ $\mu$S),  with the contributions of two sites being $f_1 = 0.41$ and $f_2 = 0.59$ respectively and the initial offset phase value $\phi=0$ throughout all the temperatures for both the sites. The presence of the Cosine oscillatory function is a classic indication of commensurate AFM ordering in a system. The fast Fourier transform (FFT) of the temporal spectrum for both temperatures is depicted in Fig. \ref{All_Final_3p05_ambient}(b). The sharpness of FFT peaks indicates the less broadening of the local field distribution which also implies the ordering to be of commensurate type. Lowering the temperature causes two FFT peaks to shift towards the higher values of the internal field. The order parameter-like behavior of the internal field as a function of temperature is shown in Fig. \ref{All_Final_3p05_ambient}(c). The phenomenological equation could be used to fit the temperature dependence of the internal magnetic field, $B_{int}(T)$ as 

\begin{figure*}
\includegraphics[scale=0.35]{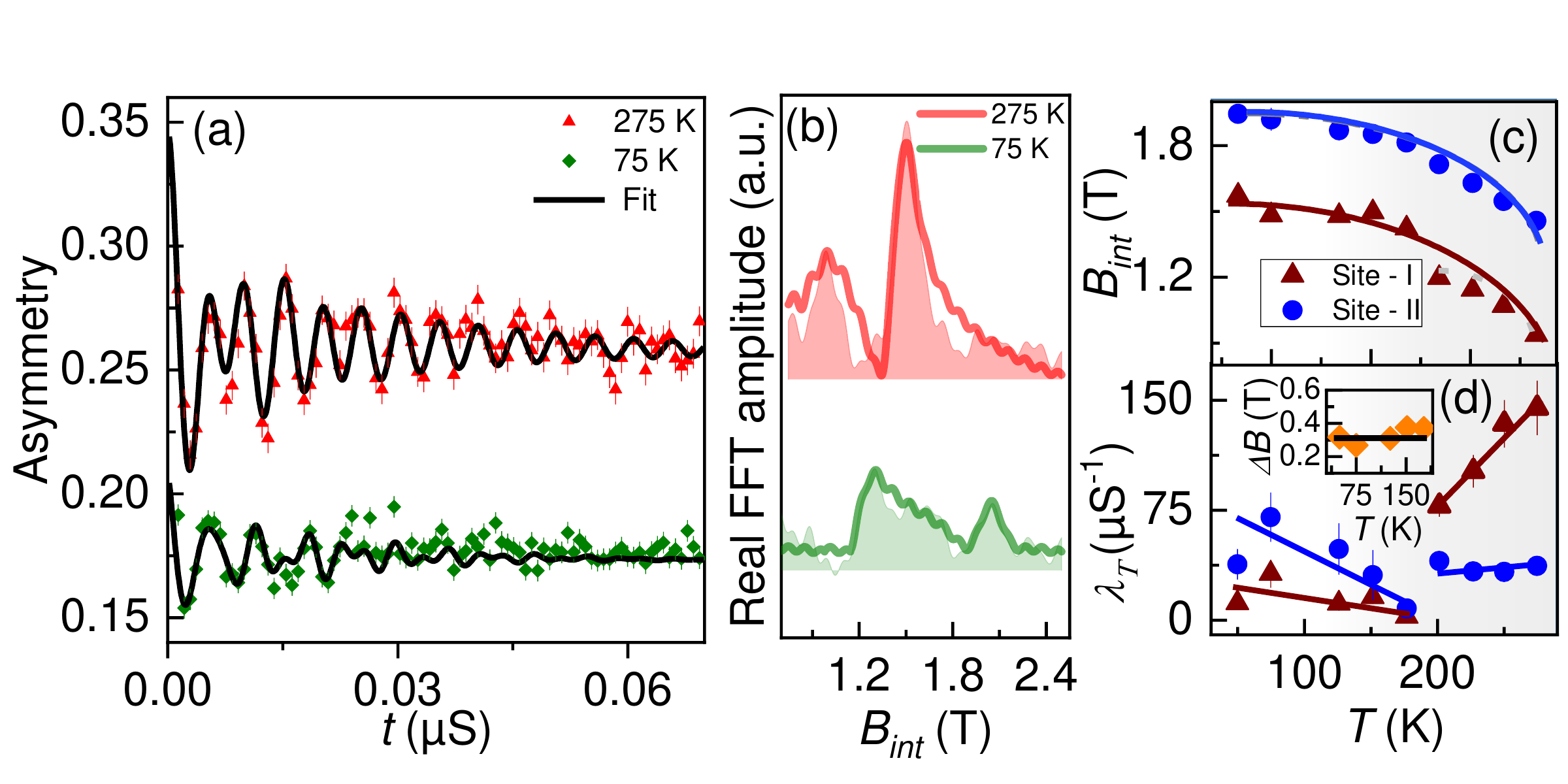}
\caption{\label{3p05_GPD_All} Mn$_{3.05}$Sn$_{0.95}$ under 1.5 GPa Pressure: \textbf{(a)} $\mu$SR asymmetry at 275~K  and 75~K (spectrum is shifted downward by 0.05 for clarity of display). Note that, the asymmetries under pressure have reduced significantly as compared to the ambient pressure condition due to the decaying relaxation of the pressure cell (see Eq. \ref{Kubo_Toyabe}). \textbf{(b)} Fourier transform amplitude of $\mu$SR asymmetry for 275~K and 75~K  where the solids lines are the theoretical depiction.  \textbf{(c)} Internal magnetic fields as a function of temperature. \textbf{(d)} Transverse relaxation rate as a function of temperature. Here solid lines are just guides to the eyes. Inset shows the value of $\Delta B$ as a function of temperature.}  
\end{figure*} 
 
\begin{equation}
\label{order_parameter}
B_{int}(T)=B_{int} (T=0)
\left[	
	1- \left( \frac{T}{T_N} \right)^\alpha
\right]^\beta
\end{equation}

where the $T_N$ is the Neel temperature and the $\alpha$, $\beta$ are two empirical parameters. The parameters were obtained by considering $T_N = 420$~K (as obtained from wTF measurement) are $\alpha=1.87(5)$, $\beta=0.65(9)$ and $B_{int,I}(T=0)=1.60(3)$~T for Site-I on the other hand $\alpha=1.99(9)$, $\beta=0.43(4)$ and $B_{int,II}(T=0)=2.01(2)$~T for Site-II. We estimated the dipolar field at two muon sites (Site-A and Site-B) and found that it is 2.01~T at Site-A and 1.60~T at Site-B. These values are in good agreement with $B_{int,II}$ and $B_{int,I}$, respectively ( for more details, see Appendix:\ref{Appendix:A}). The $\beta$-values are close to 0.5, which means the local environment of muons can be modeled using Landau mean-field theory \cite{Getzlaff2007-iu}. The $\alpha>1$ values confirm that the system exhibits some complex interactions among the magnetic moments \cite{PhysRevB.88.134416}. Because the determination of $\beta$ values is more promising in the vicinity of $T_N$ and data availability limits us to 300~K only, determining the $T_N$ based on the different $\beta$ values corresponding to the various models (e.g., three-dimensional Heisenberg or Ising model) would not be the best approach; instead, we kept the $T_N$ fixed and obtained the $\beta$ values. 

\subsubsection{$\mu^+$SR-ZF measurements under 1.5 GPa pressure}
With the application of hydrostatic pressure of 1.5 GPa, the $\mu$SR asymmetries at high temperatures (275 - 200~K) have been fitted with two Cosine functions as shown in the Eq.\ref{3p05_ambient_pressure}, which indicates that the system at higher temperature exhibits a commensurate ordering. But, for lower temperatures (175 - 50~K), Eq.\ref{3p05_ambient_pressure} was insufficient to adequately fit the $\mu$SR asymmetries. To deconvolute the asymmetries, a combination of Bessel's function and the Cosine function has been used with the site fraction of $f=0.39$, as shown in the equation

\begin{equation}
\begin{aligned}
\label{3p05_high_pressure}
	 A^{LT}_{ZF}(t)= A_0 & \biggr[ \frac{2}{3} \Bigl\{ f cos(\gamma_\mu B_1 t+\phi_1) J_0(\gamma_\mu \Delta B t) \exp^{- \frac{1}{2} (\lambda_{T_1} t)^2 }  \\	
	 				& + (1-f) J_0(\gamma_\mu B_2 		t+\phi_2)\exp^{- \frac{1}{2} (\lambda_{T_2} t)^2} \Bigl\} 
	 				 \\
	 				& +  \frac{1}{3}  \Bigl\{ f \exp^{-\lambda_{L_1}t} + (1-f) \exp^{-\lambda_{L_2}t}  \Bigl\} \biggr]
\end{aligned}
\end{equation}

Where $A_0$ is the initial asymmetry, the $J_0(\gamma_\mu \Delta B t)$ is the zeroth-order of Bessel’s function of the first kind, where $\Delta B$ is the broadening of the internal magnetic field, and $B_1$, $B_2$ are the average internal magnetic field values corresponding to the Site-I and Site-II respectively. $\phi_{1,2}$ are the initial phases and $\lambda_{T_{1,2}}$ are the Gaussian decay rates for both the sites and $\lambda_{L_{1,2}}$ is the longitudinal muon depolarization rates. As depicted in Fig.\ref{3p05_GPD_All}(a), the asymmetry of muon depolarization for 75~K fitted with the zeroth order Bessel's function of the first kind is most commonly accepted as a signature of an incommensurate (IC) field distribution at the muon sites \cite{Muon_Book_1, 20.500.11850/145189, Eyring2001-lg}.  As the Site-I is well fitted with the $cos(\gamma_\mu B_1 t+\phi_1)  J_0(\gamma_\mu \Delta B t)$, it implies that the muon stopping at Site-I experiences an incommensurate (IC) local field distribution.  The IC field distribution simply means that each muon sitting at the same crystallographic site, but experiencing different magnetic fields, with a minimum ($B_{min}$) and maximum ($B_{max}$) cut-off value, and follows a distribution function of the local magnetic field with broader distribution, reflected in the FFT spectra in Fig.\ref{3p05_GPD_All}(b) \cite{PhysRevB.65.024444}. $\Delta B $ is nothing but the range of the internal field distribution i.e., $\Delta B = B_{max} - B_{min}$, and the average of them is $B_1 = (B_{max} + B_{min}) / 2$. Muon at Site-II also experiences a broader field distribution with an average value of $B_2$ as it is well fitted with the $J_0$, which corroborates a spin-density-wave (SDW) order as approximated by the Overhauser distribution \cite{PhysRevB.84.184421, Muon_Book_1} in the limit of $B_{min} \longrightarrow 0$. 

Fig.\ref{3p05_GPD_All}(c) depicted the average internal magnetic field experienced by both the muon sites. $B_1$ and $B_2$ exhibit a generic order-parameter-like behavior as a function of temperature along with a constant value of $\Delta B$ for Site-I below T$_{IC}$ and show no anomaly near T$_{IC}$. With lowering temperature, $B_2$ is shifting towards the higher values along with the increment of $\lambda_{T_2}$, which is in contrast to the expected behavior for a commensurate ordering. This observation indicates that Site-II is also experiencing a broadening of the internal local magnetic field. These behaviors illustrates that both the sites experience a broadness of the magnetic field below T$_{IC}$ and continues throughout the low-temperature range. This is also supported by $\lambda_{T_1}$, which is almost constant below T$_{IC}$. Even though Bessel's function implies a near-zero relaxation rate in the $t \rightarrow 0$ limit, both sites have finite relaxation rates, suggesting that broadness is present at both muon-stopping sites.

The mechanism responsible for having a finite $\lambda_T$ depends on the broadness of static magnetic field distribution at muon-stopping sites. The temperature dependence of transverse relaxation rates $\lambda_T(T)$ for both the sites have been summarized in Fig. \ref{3p05_GPD_All}(d). For Site-I, $\lambda_{T_1}$ follows the typical behavior of long-range ordered AFM as the $\lambda_{T_1}$ is decreasing at temperatures $T<T_N$ \cite{Muon_Book_1}. Although it became constant below the transition temperature $T_{IC}$. For Site-II, $\lambda_{T_2}$ is almost constant within the temperature range $T_{IC}<T<T_N$, which is consistent with the commensurate AFM ordering. But, below $T_{IC}$ it increases as the temperature decreases, which implies the increasing inhomogeneity in the field distribution.
At T = 75~K the $\lambda_{T_1}=31.7 \pm 9.5$ $\mu S^{-1}$ and $\lambda_{T_2}= 70.2 \pm 16.9 ~ \mu S^{-1}$ corresponds to the half-width-half-maximum (HWHM) of local field distribution, $\Delta_1=37.980(8)$~mT and $\Delta_2(T=75~K)=82.80(5)$~mT (where $ \lambda_T = \gamma_\mu \sqrt{\Delta^2}  $) respectively gives an idea about the complex field distribution with intricate interactions with lowering the temperature \cite{PhysRevB.105.014423, PhysRevB.99.214441}. 
 
The longitudinal relaxation rate $\lambda_L$ has negligible effects on magnetic spin structure, only spin fluctuations drive the $\lambda_L$. However, one can expect that the $\lambda_L \longrightarrow 0$ when $T \rightarrow 0$ in the static limit. Since, $\lambda_L$ is very small ($<1~\mu S^{-1}$, which is not shown here), compared to the $\lambda_T$, it is difficult to conclude the dynamic nature of the system.

\subsection{Mn$_{3+x}$Sn$_{1-x}$, $x = 0.04, 0.03$}
\subsubsection{$\mu^+$SR-ZF measurements at ambient pressure}

\begin{figure}
\includegraphics[scale=0.38]{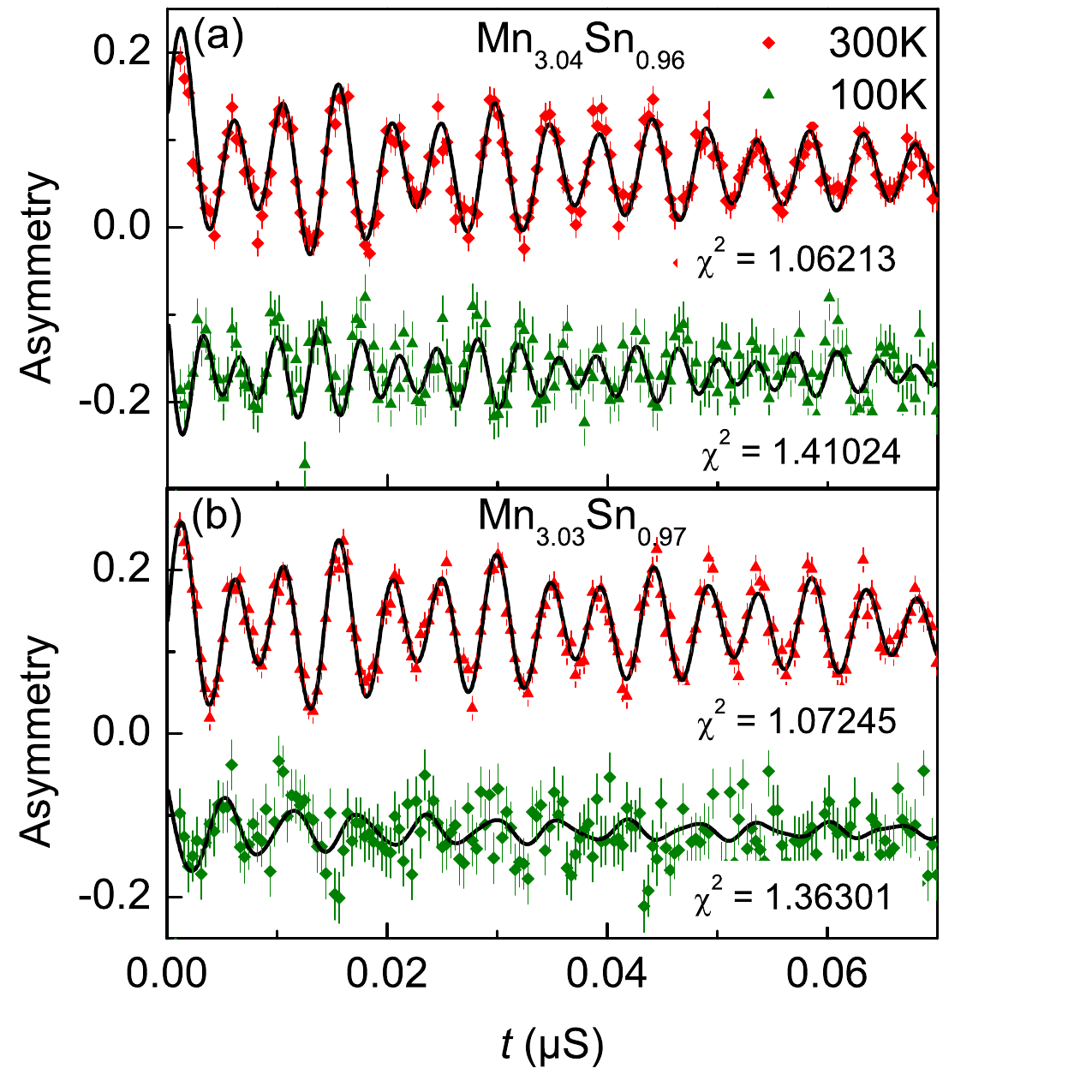}
\caption{\label{3p04+3p03_Final} Ambient Pressure $\mu$SR asymmetry at 300~K and 100~K \textbf{(a)} for Mn$_{3.04}$Sn$_{0.96}$ (spectrum is shifted downward by 0.3 for clarity of display) and \textbf{(b)} for Mn$_{3.03}$Sn$_{0.97}$ (asymmetry is shifted downward by 0.25 for clarity of display).}  
\end{figure}

For \textit{x} = 0.04 and \textit{x} = 0.03, the $\mu$SR asymmetry has been shown in Fig.\ref{3p04+3p03_Final} at two temperatures 300~K and 100~K. The best fit of the $\mu$SR asymmetry has been obtained by using Eq.\ref{3p05_ambient_pressure} for 300~K and Eq.\ref{3p03and04_equation} for 100~K.

\begin{equation}
\begin{aligned}
\label{3p03and04_equation}
	 A^{LT}_{ZF}(t)= A_0 & \biggr[ \frac{2}{3} \Bigl\{ f J_0(\gamma_\mu B_1 t) \exp^{- \frac{1}{2} (\lambda_{T_1} t)^2 }  \\	
	 				& + (1-f) cos(\gamma_\mu B_2 		t+\phi_2)\exp^{- \frac{1}{2} (\lambda_{T_2} t)^2} \Bigl\} 
	 				 \\
	 				& +  \frac{1}{3}  \Bigl\{ f \exp^{-\lambda_{L_1}t} + (1-f) \exp^{-\lambda_{L_2}t}  \Bigl\} \biggr]
\end{aligned}
\end{equation}

Cosine functions at 300~K ($T>T_{IC}$) for both sites indicate the commensurate type of AFM order. But at low temperatures, 100~K ($T<T_{IC}$) $\mu$SR asymmetry with Bessel's function at one site and a Cosine function at another site has been used, which means that the muons sitting on the Site-I, corresponding to the Bessel's function are experiencing an SDW type of order, on the other hand, the muons sitting on the Site-II are unable to experience the incommensurate nature of the spin structure. Although the data quality for 100~K is not adequate to confirm with certainty, our NPD measurements (discussed in the next section) may be pursued for a better understanding of the same.

\begin{figure}
\includegraphics[scale=0.42]{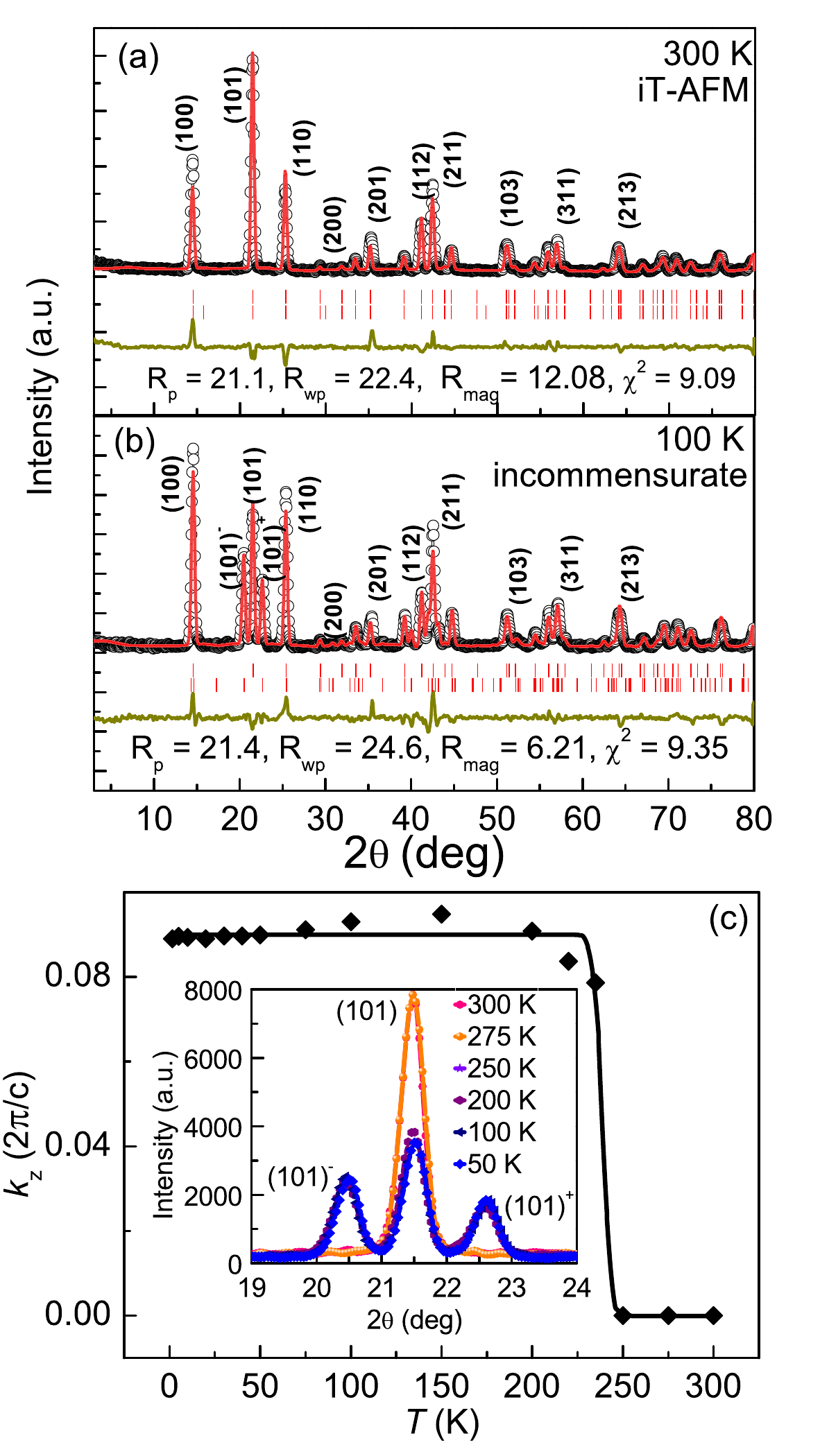}
\caption{\label{Neutron}Ambient pressure: \textbf{(a)} neutron diffraction (ND) pattern for Mn$_{3.03}$Sn$_{0.97}$ at 300~K, which is well fitted with the in-plane spin arrangement, \textbf{(b)} ND pattern at 100~K fitted with the up-down canted model.  \textbf{(c)} modulation wave-vector $k=(0,0,k_z)$ of the helical spin arrangement. Inset of \textbf{(c)} depicts zoomed-in NPD spectra at different temperatures around (101) peak, which shows two satellite peaks $(101)^\pm$ are arising below 250~K.}  
\end{figure}

\subsubsection{Neutron diffraction at ambient pressure on Mn$_{3.03}$Sn$_{0.97}$}
The Rietveld refinement of the NPD pattern for the temperature of 300~K is shown in Figure \ref{Neutron}(a). This confirms that the room temperature magnetic structure can be nicely fitted to an in-plane anti-chiral $(\chi=-1)$ inverse triangular spin structure, as has been widely reported in the literature \cite{wang2023flat, P_J_Brown_1990}. The magnetic moment of Mn-atoms is calculated as 2.36~$\mu_B$. By lowering the temperature, it is observed that well-defined satellite peaks start to appear at temperatures below 250~K. The presence of satellite peaks $(101)^\pm$ indicates the existence of a modulated magnetic structure in this sample. Our Rietveld refinements establish the presence of a helical modulation of the inverse triangular spin structure along the c-axis, with a modulation vector $\textbf{k} = (0, 0, 0.089)$. The Rietveld refinement of  NPD data for a temperature of 100~K is shown in Figure \ref{Neutron}(b). Some of the peak intensity can be accounted for by using a modulation of the up-down canted model for low-temperature data, as discussed in Ref.~\cite{PhysRevB.101.144422}. Figure \ref{Neutron}(c) depicts the modulation wave vector along the z-direction ($k_z$) as a function of temperature. This is also in line with the \textbf{k}-value (in the longitudinal direction of single crystal Mn$_3$Sn) of 0.084 previously reported in the Ref.~\cite{PhysRevResearch.6.L032016}. 

\begin{figure*}[btp]
\includegraphics[scale=0.55]{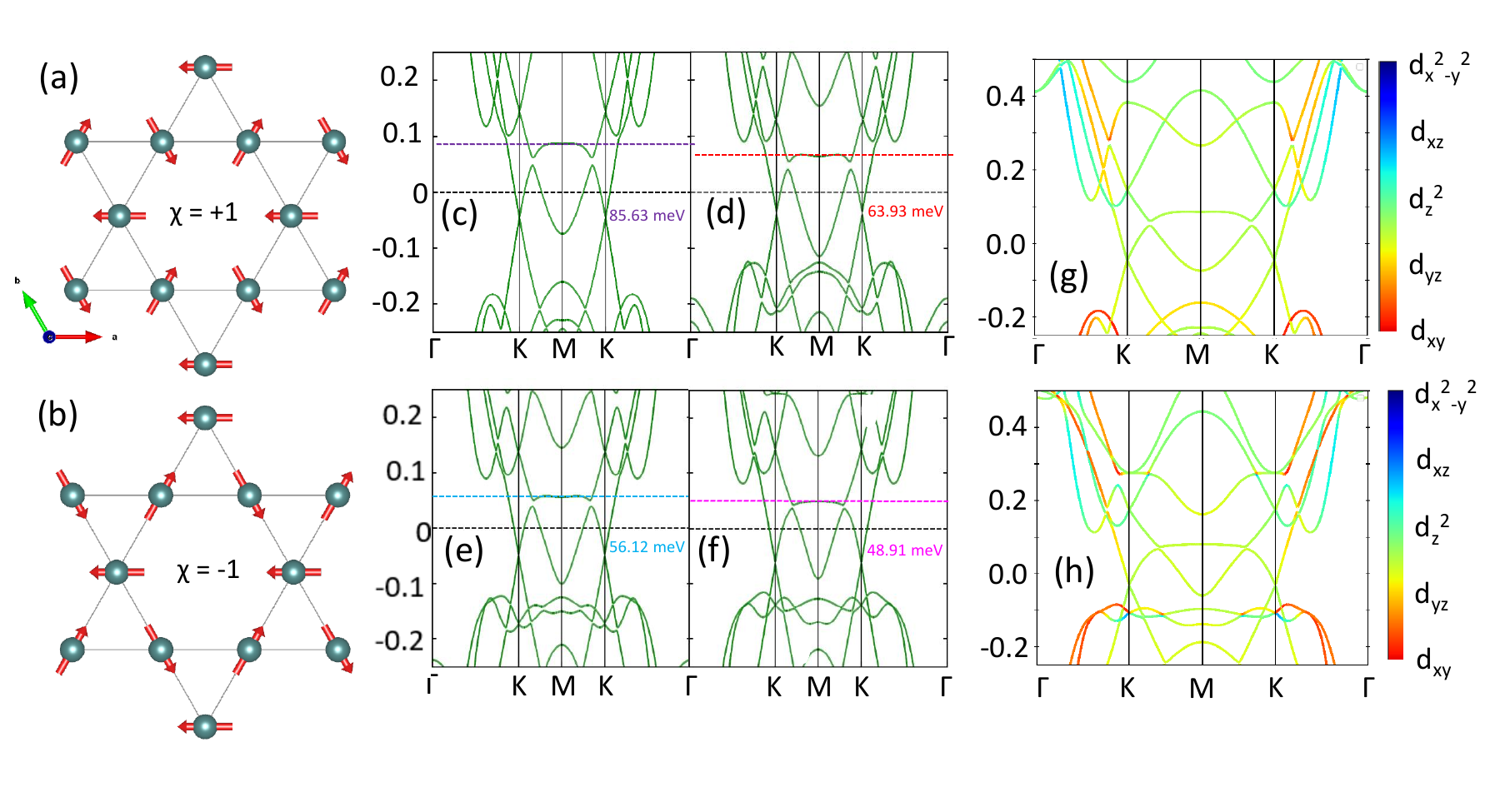}
\caption{\label{BS}\textbf{(a)} Chiral $\chi=+1$ \textbf{(b)} anti-chiral $\chi=-1$ spin-arrangement of Mn-atoms. \textbf{(c-f)} Calculated $GGA+SOC$ band structures for Mn$_3$Sn in $\chi=-1$ configuration under 0 GPa, 0.64 GPa, 1.29 GPa, and 1.93 GPa pressures respectively. The black dashed line represents the Fermi energy for the respective structures (set to 0 eV). The purple, red, blue, and pink dashed lines are marked as the energy level of the flat band for the 0 GPa to 1.93 GPa pressure cases respectively. \textbf{(g-h)} Orbital-projected band structure for 0 GPa and 1.93 GPa respectively.}  
\end{figure*}

\begin{figure*}[btp]
\includegraphics[scale=0.60]{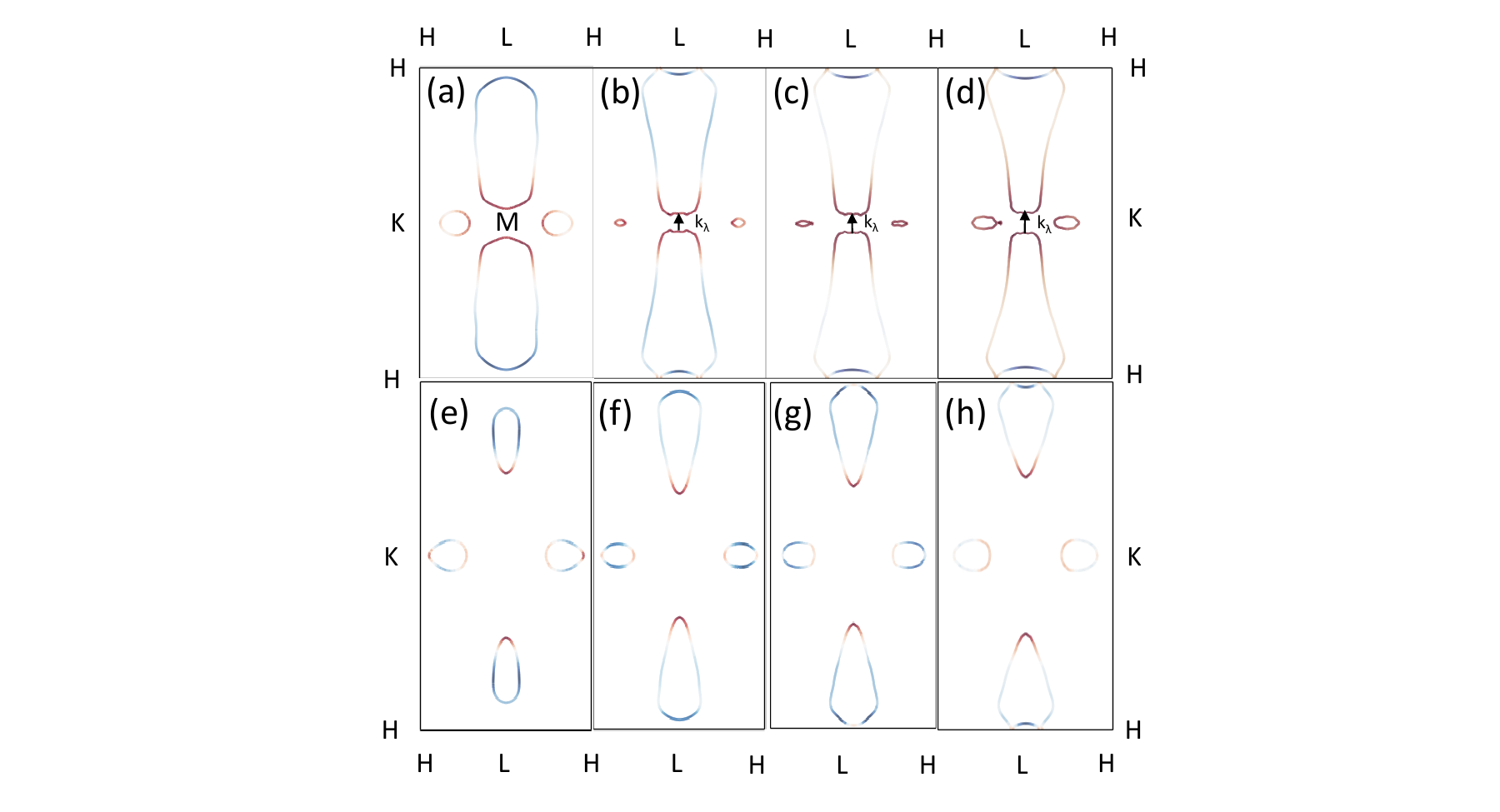}
\caption{\label{FS} \textbf{(a - d)} Fermi surfaces plotted across the [100] plane under 0 GPa, 0.64 GPa, 1.29 GPa, and 1.93 GPa pressure respectively at the flat band energy level chemical potential. The nesting vectors are $k_{\lambda}$=0.1252 ,$k_{\lambda}$=0.0782, $k_{\lambda}$=0.0789, $k_{\lambda}$=0.0888 for 0 GPa, 0.64 GPa, 1.29 GPa, and 1.93 GPa pressure respectively. \textbf{(e - h)} Fermi surfaces plotted across the [100] plane under 0 GPa, 0.64 GPa, 1.29 GPa, and 1.93 GPa pressure respectively at the Fermi energy.}  
\end{figure*}

\section{DISCUSSION}
To comprehend the consequence of hydrostatic and chemical pressure on Mn$_3$Sn, we'll take a closer look at the bulk band structure, especially the topological flat band features, typically present in kagome structures. Flat bands closer to the Fermi level are more likely to meet the Fermi-nesting condition by having a characteristic wave vector (\textit{k}) associated with CDW and SDW instability, which can lead to Van Hove singularities \cite{PhysRevB.107.205130, Okamoto2022, Kang2020}. Our bulk band structure calculations on Mn$_3$Sn suggest that the application of hydrostatic pressure brings the flat band closer to the Fermi level, increasing the chance of exhibiting the Fermi nesting condition as illustrated in Fig. \ref{BS}(c-f). The Fermi surfaces as shown in Fig. \ref{FS}(a-h), further demonstrate that the area of the nested surfaces associated with the Fermi nesting vector (k) increases with pressure, which favors stabilizing a complex ordering. Thus, the shifting of flat bands and Weyl nodes near the Fermi level in the lower temperature region manifests itself as a magnetic state change from the iT-AFM to the helical phase. This is supported by the neutron and $\mu^+$SR data analysis.

A recent study on single crystal Mn$_{3+x}$Sn$_{1-x}$ ($x=0.012$) also confirms the presence of incommensurate non-coplanar spin structure, reflecting complex inter-band nature of the nesting with the longitudinally polarized SDW (with wavevector $k_\lambda$) and transversely polarized helimagnet (with wavevector $k_\tau$), as well as CDW type of order with similar k-values of our calculated k-values \cite{ wang2023flat}. Since the hexagonal structure of Mn$_3$Sn only stabilizes in the off-stochiometric composition with higher electronic concentration due to the excess Mn replacing some of the Sn atoms as reported in earlier studies \cite{Zhang_2013, KREN1975226, doi:10.1143/JPSJ.51.2478}, extra electrons contributed by excess Mn have an important role in the bulk band structure. This excess electron concentration may lift the chemical potential level towards the flat band. This is understood as the shifting of the Fermi level near the flat band is only achievable with weakly doped concentrations ($x=0.03, 0.04$) because over-doping ($x=0.05$) might push the Fermi level far away from the flat band resulting in a lower possibility of fulfilling the criteria for Fermi-surface nesting \cite{wang2023flat}. This might explain the fact that no incommensurate transition was found in Mn$_{3.05}$Sn$_{0.95}$, near $T_{IC}$ at ambient pressure. However for Mn$_{3.05}$Sn$_{0.95}$ the application of hydrostatic pressure modifies its bond lengths between the Mn-triangle, which is directly related to the exchange parameters between the Mn-atoms \cite{Nayak2020}, hence, the alteration of the magnetic band structure. This causes the Fermi level to be shifted towards the flat band, resulting in an IC-complex type of ordering as apparent from the ZF-$\mu^+$SR analysis under pressure. 

Although both physical hydrostatic pressures and doping exert positive pressure on the system, and the magnetic ground state behaves similarly in low-temperature regions, they have selected a somewhat different mechanism to induce the IC-complex form of order. X-ray diffraction (XRD) study revealed that the lattice parameters were changed by pressure about 150 times more than by doping. This suggests that doping was not as effective as pressure in changing bond length \cite{Nayak2020}. In summary, on the one hand, a slight excess electron concentration causes the Fermi level in the vicinity of the flat band, hence, inducing IC-complex ordering without significantly changing the bond length. On the other hand, hydrostatic pressure modifies magnetic exchanges, hence, magnetic band structure and this brings the Fermi level from the far-off position caused by over-doping towards the flat band, introducing the incommensurate CDW and SDW kind of order. Although a small change in the band topology may lead to vanishing AHE this can be understood as bands amalgamating with CDW and SDW. However, a small doping of x = 0.03, 0.04, and 0.05, equal to 1\%, 1.3\%, and 1.67 \% of excess Mn in Mn$_3$Sn, respectively, have a high impact on physical properties, which is extremely unusual; further research is required to understand the explanation for such a response of this compound to doping.

\section{conclusion}
We have successfully employed two powerful microscopic techniques such as $\mu$SR and neutron diffraction to study the evolution of magnetic states induced by hydrostatic pressure and doping (by varying $x$) in Mn$_{3+x}$Sn$_{1-x}$. Also using band structure calculation we discuss the possible reason for the commensurate to incommensurate transition at a lower temperature. $\mu^+$SR experiments on Mn$_{3.05}$Sn$_{0.95}$ reveal a commensurate magnetic state throughout all temperatures below the Neel ordering $T_N \approx 420$~K under ambient pressure. We have observed two muon sites experimentally which was also supported by our \textit{ab-initio} calculations. Dipolar field calculations also match with the experimentally obtained local field at the muon sites considering iT-AFM. Interestingly, the application of 1.5~GPa hydrostatic pressure introduces an additional incommensurate magnetic ordering below $T_{IC} \sim 175$~K. A similar behavior of commensurate to incommensurate transition is also observed with the composition of $x = 0.04$ and $x=0.03$ using the $\mu$SR. Elastic neutron scattering technique is employed on the $x=0.03$ compound which further supports the presence of incommensurate ordering with a modulation vector below  $T_{IC}$ is $\textbf{k} = (0, 0, 0.089)$. Moreover, we have discussed the effect and reason for the development of an incommensurate magnetic ordering at low temperatures due to Fermi nesting under hydrostatic or chemical pressure based on our band-structure calculations.

Nonetheless, the microscopic evidence that incommensurate spin spiral ground state stabilizes by applying hydrostatic pressure or changing the Mn concentration is now established from this work. This might help us to understand the behavior of the large AHE at different temperatures via band topology in a more sophisticated manner. Alternatively, instead of using traditional FMs, more studies might be done on how the AHE responds to pressure and soft magnetic fields to learn more about its prospects for use in-memory storage and spintronics devices.

\appendix 
\section{Computational Details \label{Appendix:A}}
First principles electronic structure calculations have been carried out to understand the microscopic impact of hydrostatic pressure on the underlying electronic and the magnetic state of the material. The density functional theory-based electronic structure calculations are carried out in the plane wave basis set with projector augmented wave (PAW) pseudopotential as implemented in the Vienna ab initio Simulation Package (VASP) \cite{PhysRevB.54.11169,PhysRevB.59.1758}. The generalized gradient approximation (GGA) based exchange-correlation is used in the form of Perdew-Burke-Ernzerhof (PBE) functional \cite{PhysRevLett.77.3865}. The 8$\times$8$\times$9 $\Gamma$- centered $k$-point mesh is used for the Brillouin zone (BZ) integration with an energy cutoff of 400 eV and an energy convergence threshold of $10^{-7}$ eV used to get a converged charged density. The structural relaxations have been done until the Hellmann-Feynmann force on each atom becomes less than 0.01 eV/$\AA$. The effect of the SOC has been taken into account as a relativistic correction to the original Hamiltonian\cite{PhysRevB.62.11556}. The Fermi surface is plotted with $\Gamma$ -centered K-point mesh with a resolution of 0.01 $\AA$, using ifermi\cite{Ganose2021}. 

\begin{figure}[btp]
\includegraphics[scale=0.25]{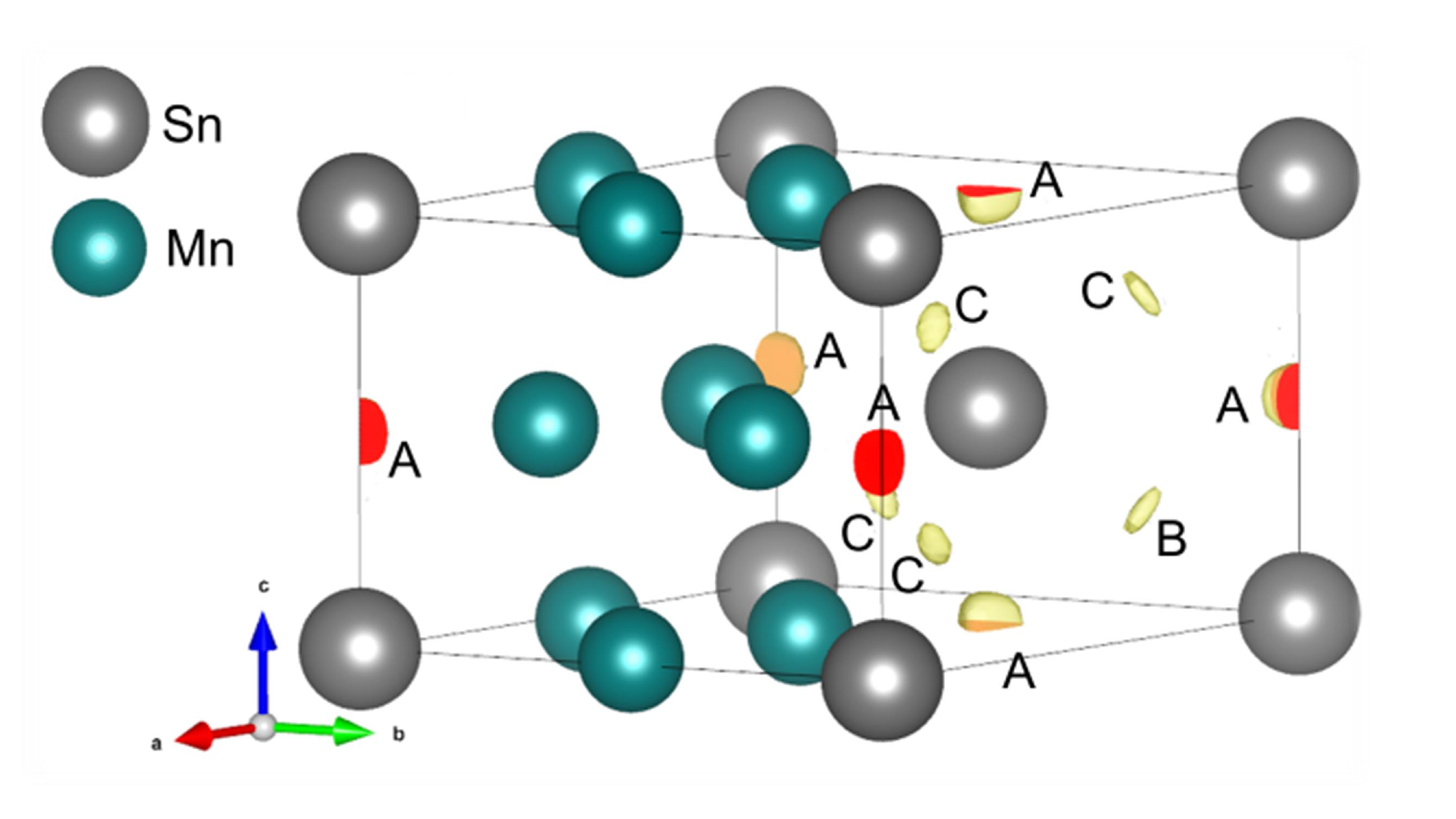}
\caption{\label{DFT_isosufaces} Local potential isosurfaces in a unit cell of Mn$_3$Sn (note that the unit cell only consists 6 Mn-atoms and 2 Sn-atoms, here other than these 8 atoms are from nearest unit cell) with isosurface level in positive iso-surface values of 1 eV (i.e A-type), for 0.8 eV (i.e B and C-types) have emerged.}  
\end{figure}

\subsection{Band structure calculations}
In kagome lattice spins in a 120$^o$ structure may arrange among themselves in two modes i.e chiral ($\chi$=+1) or anti-chiral ($\chi$=-1) configurations as shown in Fig. \ref{BS}(a) and (b) respectively. The total energy calculations show that the anti-chiral configuration is energetically lower by 2.5 meV/f.u compared to the chiral arrangement, which is further consistent with NPD results. Therefore rest of the electronic structure calculations are done with this on the $\chi=-1$ AFM state. The calculated magnetic moment 3$\mu_B$/Mn site, which is consistent with the previously reported results \cite{P_J_Brown_1990, Tomiyoshi_Yamaguchi_1982,PhysRevB.101.144422, wang2023flat, PhysRevResearch.6.L032016}.

The effect of applied uniform hydrostatic pressure is replicated by contracting the lattice parameters by 1, 2, and 3 $\%$ concerning the equilibrium lattice parameters. We have calculated the corresponding hydrostatic pressure using the bulk modulus (64.5 GPa) of Mn$_3$Sn \citep{PhysRevB.105.174430}, and turn out to be the pressures of 0.645 GPa, 1.29 GPa, and 1.93 GPa respectively for the above-mentioned reduction of the lattice parameters. The calculated GGA+SOC band structures for the ambient and high-pressure structures are shown in Fig. \ref{BS}(c-f). The calculated band structures as shown in Fig. \ref{BS}(c) are similar to those reported in the previous works \cite{wang2023flat}, particularly the most striking feature of the band structure is the flat band along the high-symmetry $K-M-K$ direction in the BZ. The previous reports of the electron-doped Mn$_3$Sn clearly show the upward shifting of the Fermi energy closer to the flat band. In our calculations with increasing pressure, the flat band moved downward closer to the Fermi energy as marked by the dashed lines in Fig. \ref{BS}(c-f). We also have quantified the shifting of the flat band and found that with increasing pressure from 0 to 1.93 GPa, the separation between the flat band and the Fermi level reduced by nearly half i.e., 85.63 to 48.91 meV. An interesting point to be noted is that the application of pressure up to 1.93 GPa, does not affect the band structure significantly at the Fermi level and the linear crossing at the $K$-point remains intact in the high-pressure cases as well. To understand the band structure in detail, we have calculated the orbital projected band structure of the ambient pressure (i.e. 0 GPa) and high pressure (i.e. 1.93 GPa) cases as shown in Fig. \ref{BS}(g-h). Orbital contribution indicates that close to Fermi energy the dominating nature of the bands are of $d_{3z^2}$ character. The flat bands along the $K-M-K$ path are also majorly  $d_{3z^2}$ character with a small admixer of the $d_{yz}$ character for the 0 GPa case shown in Fig. \ref{BS}(g). However, in the 1.93 GPa high-pressure case shown in Fig. \ref{BS}(h), the band character minorly changes, where the contribution of the $d_{yz}$ character in the flat band and the linear dispersion increases. Point to note that, with the increasing pressure, the orbital character at the Fermi energy and the flat band remains intact and both bands are dominantly of $d_{3z^2}$ character.

With the application of pressure although the orbital character hardly changed, however, the band dispersion has noticeable changes. Apart from the shifting of the flat band close to the Fermi energy, the flat band dispersion along the $K-M-K$ path has modified with the pressure. To understand in detail the band dispersion in the BZ we have calculated the Fermi surface for the [100] hexagonal plane at the chemical potential equal to that of the flat band energy and the Fermi energy, as shown in Fig.\ref{FS}(a-d) and Fig.\ref{FS}(e-h) respectively for four different pressures. Comparing the Fermi surfaces, it is clear that with increasing pressure, the nature of the pockets in the BZ has changed, which is consistent with the band dispersion and the system is becoming more prone to exhibit a Fermi nesting condition, where the flat band (shown in Fig.\ref{FS}(a-d)) leads to a possible Fermi nesting condition which leads the system to CDW or SDW or coexistence of both. A close look at the pockets near the $M$ points becomes more narrow in the case of 1.93 GPa pressure (Fig.\ref{FS}(d)) compared to that of the 0 GPa case (Fig.\ref{FS}(a)), where the pocket close to circle. Therefore, one can conclude that with increasing pressure, the flat bands approaching the Fermi energy with become more flatter. $k$-value ( k$_{\lambda}$ = 0.089) that we found from the Fermi surface for 1.93 GPa hydrostatic pressure at the flat band level is in line with the experimental $k$-value found from our NPD analysis, which is also in agreement with recent neutron studies on single crystal Mn$_3$Sn \cite{PhysRevB.101.144422, PhysRevResearch.6.L032016}. And the k$_{\lambda}$ is consistent with the k-value obtained from the NPD analysis. As the pressure increases, $k$-values decrease, and the surface area covered in the BZ has enhanced. From Fig.\ref{FS}(a-d), it is evident that, with increasing pressure larger areas of the fermi surface associated with the nesting vectors fulfill the nesting condition in line with our earlier experimental predictions. Fig.\ref{FS}(e-h) represents the Fermi surfaces at the Fermi energy for increasing pressures respectively. Although there are no drastic changes in the Fermi pockets locations in the BZ, the Fermi surface area has increased with the increase of pressure. From Fermi surface analysis, it is evident that with increasing pressure larger areas of the Fermi surface associated with the nesting vectors fulfill the nesting condition in line with our earlier experimental predictions.


\begin{table*}
\large
\begin{center}
\caption{\label{muon_site_table}The calculated positions and energies for different muon sites and their corresponding dipolar magnetic field compared with the experimentally measured B$_{int}$.}
\begin{tabular}{ |c||c|c|c|c|c|c } 
\hline
Site & Site position & Energy (eV) & Site type & Cal. B$_{int}$ (T) & Exp. B$_{int} $~(T) \\
\hline
\hline
1 & 0.000, 0.000, 0.500 & -243.769 & \multirow{3}{1em}{A} & 2.073 & $B_{int,II}=2.01$ \\ 
2 & 0.333, 0.667, 0.008 & -243.765 &  & 2.069 &  \\ 
3 & 0.333, 0.667, 0.992 & -243.765 &  & 2.068 &  \\
\hline
4 & 0.166, 0.836, 0.252  & -243.891 & B & 1.4033 & $B_{int,I}=1.60$ \\
\hline
5 & 0.166, 0.333, 0.749  & -241.389 & \multirow{5}{1em}{C} & 0.7183 & \multirow{5}{1em}{-} \\ 

6 & 0.166, 0.333, 0.250  & -243.070 &  & 0.7183 & \\
7 & 0.166, 0.835, 0.749  & -241.397 &  & 1.4033 & \\  
8 & 0.667, 0.833, 0.249  & -241.000 &  & 1.4033 & \\ 
9 & 0.667, 0.833, 0.750  & -241.394 &  & 1.4033 & \\  
\hline
\end{tabular}
\end{center}
\begin{minipage}{16.5cm}
\small{$\blacktriangleright$ Note: Here Cal. $B_{int}$ is the estimated dipolar magnetic field calculated theoretically and Exp. $B_{int}$ is the internal magnetic field obtained from the $\mu^+$SR data analysis.}
\end{minipage}

\end{table*}


\subsection{Muon sites calculations}
To estimate the muon sites in Mn$_{3+x}$Sn$_{1-x}$ unit cell, we have performed the DFT calculation with stoichiometric Mn$_3$Sn to reduce the computational cost. The calculations were performed using $5\times 5 \times 11$ $\Gamma$-centered $K$-mesh. A pseudopotential of positively-charged hydrogen (H$^+$) was used to mimic the $\mu^+$. We primarily focused on finding out the regions that have high electron density, where the $\mu^+$ could be placed. To do this we calculated the local potentials for Mn$_3$Sn with $2\times2\times1$ supercell and analyzed the isosurfaces to identify the possible muon sites. The positive isosurfaces at local potential at 1 eV give two different positions for possible muon sites, which we denoted here as A-type shown in Fig. \ref{DFT_isosufaces}, whereas by decreasing the value of the isosurface level to 0.8 eV, another two new types of positions emerge in the cell at the 6h Wyckoff position, labeled as B and C-type respectively as shown in Fig. \ref{DFT_isosufaces}. It is visible that the iso-surface size of the A-type is significantly larger than the B/C-type, which indicates that the strength of the negative potential at the A-site is larger than the B/C-sites. Muons were inserted at 9 different possible muon sites, relaxed, and the total energies of the supercells were calculated self-consistently. These calculations show that the B-type is the energetically lowest followed by A-type sites, which are marginally higher in energy ($\Delta E \sim$~122 meV/f.u). The C-type sites have different energies but all of them have higher energy than that of the A and B types. The coordinates of the muon sites and the corresponding energies are tabulated in TABLE \ref{muon_site_table}.

\subsection{Dipolar field calculatios}
The internal magnetic field ($B_{int}$) encountered by $\mu^+$ at the muon site is generally a combined effect of dipolar field ($B_{dip}$), contact field ($B_{cont}$) and the Lorentz field ($B_L$) as shown in below equation \cite{Muon_Book_1}

\begin{equation}
\label{magnetic_field}
B_{int}=B_{dip}+B_{cont}+B_L
\end{equation}

In general $B_{int}$ is dominated by the $B_{dip}$ as $B_{cont}$ and $B_L$ (for A-type antiferromagnets $B_L=0$) has less contributions in case of $\mu^+$SR. The $B_{dip}$ can be expressed as 

\begin{equation}
\label{dipolar_field}
B_{dip}=\frac{\mu_0}{4 \pi} \sum_{i=1}^N \left(
 -\frac{m_i}{r_i^3} + \frac{3(m_i.r_i)r_i}{r_i^5}
 \right) 
\end{equation}
symbols used here have their typical meanings.

We calculated the dipolar magnetic field for each of the muon sites estimated from the DFT simulations using the Magnetic structure and mUon Embedding Site Refinement (MuESR) package \cite{doi:10.7566/JPSCP.21.011052}, for $\chi=-1$ AFM with magnetic space group \textit{Cmc'm'} as shown in Fig.\ref{BS}(b). As the refined value of the magnetic moment of individual Mn-atoms depends on their synthesis process and stochiometric values, different kinds of literature reported different values for the same \cite{P_J_Brown_1990, Tomiyoshi_Yamaguchi_1982,PhysRevB.101.144422, wang2023flat, PhysRevResearch.6.L032016}, here we have used the value  $m = 2.54 \mu_B$ as reported in Ref.\cite{wang2023flat}. Three different values of $B_{dip}$ have been estimated as 2.073, 1.4033, and 0.7183~T. Since the type-C sites are energetically high and less probable to attract $\mu^+$, we have considered type-A and type-B sites only. The estimated $B_{dip}$ is calculated using crystallographic information of ideal stoichiometry. 

Nevertheless, $B_{dip}$ for Site-A is close enough to the experimental value of 2.01~T and for B is close to 1.6~T. However C-type muon sites are unable to be observed with our data, they could be noticeable with high-resolution data obtained from a single crystal. The calculated dipolar field for the muon site is consistent with the experimental results from $\mu^+$SR analysis.  Thus, our computational findings confirm the existence of two possible muon sites in Mn$_3$Sn and provide additional backing for our $\mu^+$SR data analysis.

\section{Pressure cell Contribution}
In the measurement process under pressure, muons stopping in the pressure cell surrounding the sample reduces a significant portion of the $\mu$SR asymmetry. As a result, the $\mu$SR data across the whole temperature range were analyzed by decomposing the signal into a sample contribution and a pressure cell contribution as

\begin{equation}
\label{pressure_cell_contribution}
A(t)=A_0 \bigr[ FP_{PC}(t) + (1-F)P_S(t) \bigr]
\end{equation}

where $A_0$ is the initial asymmetry, F is the fraction sharing between the sample and the pressure cell and $P_S(t)$ and $P_{PC}(t)$ are the time-dependent polarization of muon stopping in the sample and pressure respectively. The pressure contribution exhibits a damped Kubo-Toyabe function \cite{kubo1967magnetic} with an exponential term,

\begin{equation}
\label{Kubo_Toyabe}
P_{PC}(t) = \biggr[ \frac{1}{3}+\frac{2}{3} \Bigl\{1-(\sigma t)^2  \Bigl\} e^{-\frac{1}{2}(\sigma t)^2}  \biggr] e^{-\lambda t}
\end{equation}

Where the $\sigma/\gamma_{\mu}$ is the width of the local field distribution and $\lambda$ is the electronic relaxation rate. The temperature dependence of these parameters is well documented in Ref.\cite{doi:10.1080/08957959.2017.1373773, doi:10.1080/08957959.2016.1173690}. The sample contribution is nothing but the same function we have used throughout the discussion i.e., $A_{ZF}(t)=A_0 P_S(t)$. In our case, the contribution from the pressure cell was about 50 percent.

\section*{ACKNOWLEDGMENTS}
We gratefully appreciate the assistance offered by Prof. Priya Johari, SNIoE, Delhi NCR, India, during the preliminary DFT calculations to determine the muon sites. We are also grateful to Dr. Pietro Bonfa, University of Parma, Italy for his assistance in configuring the dipolar field computational setup. We also thank Dr. Alex Ganose, Imperial College London for his assistance with ifermi\citep{Ganose2021}.

\bibliography{apssamp}

\providecommand{\noopsort}[1]{}\providecommand{\singleletter}[1]{#1}%
\begin{thebibliography}{70}%
\makeatletter
\providecommand \@ifxundefined [1]{%
 \@ifx{#1\undefined}
}%
\providecommand \@ifnum [1]{%
 \ifnum #1\expandafter \@firstoftwo
 \else \expandafter \@secondoftwo
 \fi
}%
\providecommand \@ifx [1]{%
 \ifx #1\expandafter \@firstoftwo
 \else \expandafter \@secondoftwo
 \fi
}%
\providecommand \natexlab [1]{#1}%
\providecommand \enquote  [1]{``#1''}%
\providecommand \bibnamefont  [1]{#1}%
\providecommand \bibfnamefont [1]{#1}%
\providecommand \citenamefont [1]{#1}%
\providecommand \href@noop [0]{\@secondoftwo}%
\providecommand \href [0]{\begingroup \@sanitize@url \@href}%
\providecommand \@href[1]{\@@startlink{#1}\@@href}%
\providecommand \@@href[1]{\endgroup#1\@@endlink}%
\providecommand \@sanitize@url [0]{\catcode `\\12\catcode `\$12\catcode
  `\&12\catcode `\#12\catcode `\^12\catcode `\_12\catcode `\%12\relax}%
\providecommand \@@startlink[1]{}%
\providecommand \@@endlink[0]{}%
\providecommand \url  [0]{\begingroup\@sanitize@url \@url }%
\providecommand \@url [1]{\endgroup\@href {#1}{\urlprefix }}%
\providecommand \urlprefix  [0]{URL }%
\providecommand \Eprint [0]{\href }%
\providecommand \doibase [0]{https://doi.org/}%
\providecommand \selectlanguage [0]{\@gobble}%
\providecommand \bibinfo  [0]{\@secondoftwo}%
\providecommand \bibfield  [0]{\@secondoftwo}%
\providecommand \translation [1]{[#1]}%
\providecommand \BibitemOpen [0]{}%
\providecommand \bibitemStop [0]{}%
\providecommand \bibitemNoStop [0]{.\EOS\space}%
\providecommand \EOS [0]{\spacefactor3000\relax}%
\providecommand \BibitemShut  [1]{\csname bibitem#1\endcsname}%
\let\auto@bib@innerbib\@empty
\bibitem [{\citenamefont {Lv}\ \emph {et~al.}(2015)\citenamefont {Lv},
  \citenamefont {Weng}, \citenamefont {Fu}, \citenamefont {Wang}, \citenamefont
  {Miao}, \citenamefont {Ma}, \citenamefont {Richard}, \citenamefont {Huang},
  \citenamefont {Zhao}, \citenamefont {Chen}, \citenamefont {Fang},
  \citenamefont {Dai}, \citenamefont {Qian},\ and\ \citenamefont
  {Ding}}]{PhysRevX.5.031013}%
  \BibitemOpen
  \bibfield  {author} {\bibinfo {author} {\bibfnamefont {B.~Q.}\ \bibnamefont
  {Lv}}, \bibinfo {author} {\bibfnamefont {H.~M.}\ \bibnamefont {Weng}},
  \bibinfo {author} {\bibfnamefont {B.~B.}\ \bibnamefont {Fu}}, \bibinfo
  {author} {\bibfnamefont {X.~P.}\ \bibnamefont {Wang}}, \bibinfo {author}
  {\bibfnamefont {H.}~\bibnamefont {Miao}}, \bibinfo {author} {\bibfnamefont
  {J.}~\bibnamefont {Ma}}, \bibinfo {author} {\bibfnamefont {P.}~\bibnamefont
  {Richard}}, \bibinfo {author} {\bibfnamefont {X.~C.}\ \bibnamefont {Huang}},
  \bibinfo {author} {\bibfnamefont {L.~X.}\ \bibnamefont {Zhao}}, \bibinfo
  {author} {\bibfnamefont {G.~F.}\ \bibnamefont {Chen}}, \bibinfo {author}
  {\bibfnamefont {Z.}~\bibnamefont {Fang}}, \bibinfo {author} {\bibfnamefont
  {X.}~\bibnamefont {Dai}}, \bibinfo {author} {\bibfnamefont {T.}~\bibnamefont
  {Qian}},\ and\ \bibinfo {author} {\bibfnamefont {H.}~\bibnamefont {Ding}},\
  }\bibfield  {title} {\bibinfo {title} {Experimental discovery of weyl
  semimetal \uppercase{T}a\uppercase{A}s},\ }\href
  {https://doi.org/10.1103/PhysRevX.5.031013} {\bibfield  {journal} {\bibinfo
  {journal} {Phys. Rev. X}\ }\textbf {\bibinfo {volume} {5}},\ \bibinfo {pages}
  {031013} (\bibinfo {year} {2015})}\BibitemShut {NoStop}%
\bibitem [{\citenamefont {Xu}\ \emph {et~al.}(2015)\citenamefont {Xu},
  \citenamefont {Belopolski}, \citenamefont {Alidoust}, \citenamefont
  {Neupane}, \citenamefont {Bian}, \citenamefont {Zhang}, \citenamefont
  {Sankar}, \citenamefont {Chang}, \citenamefont {Yuan}, \citenamefont {Lee},
  \citenamefont {Huang}, \citenamefont {Zheng}, \citenamefont {Ma},
  \citenamefont {Sanchez}, \citenamefont {Wang}, \citenamefont {Bansil},
  \citenamefont {Chou}, \citenamefont {Shibayev}, \citenamefont {Lin},
  \citenamefont {Jia},\ and\ \citenamefont
  {Hasan}}]{doi:10.1126/science.aaa9297}%
  \BibitemOpen
  \bibfield  {author} {\bibinfo {author} {\bibfnamefont {S.-Y.}\ \bibnamefont
  {Xu}}, \bibinfo {author} {\bibfnamefont {I.}~\bibnamefont {Belopolski}},
  \bibinfo {author} {\bibfnamefont {N.}~\bibnamefont {Alidoust}}, \bibinfo
  {author} {\bibfnamefont {M.}~\bibnamefont {Neupane}}, \bibinfo {author}
  {\bibfnamefont {G.}~\bibnamefont {Bian}}, \bibinfo {author} {\bibfnamefont
  {C.}~\bibnamefont {Zhang}}, \bibinfo {author} {\bibfnamefont
  {R.}~\bibnamefont {Sankar}}, \bibinfo {author} {\bibfnamefont
  {G.}~\bibnamefont {Chang}}, \bibinfo {author} {\bibfnamefont
  {Z.}~\bibnamefont {Yuan}}, \bibinfo {author} {\bibfnamefont {C.-C.}\
  \bibnamefont {Lee}}, \bibinfo {author} {\bibfnamefont {S.-M.}\ \bibnamefont
  {Huang}}, \bibinfo {author} {\bibfnamefont {H.}~\bibnamefont {Zheng}},
  \bibinfo {author} {\bibfnamefont {J.}~\bibnamefont {Ma}}, \bibinfo {author}
  {\bibfnamefont {D.~S.}\ \bibnamefont {Sanchez}}, \bibinfo {author}
  {\bibfnamefont {B.}~\bibnamefont {Wang}}, \bibinfo {author} {\bibfnamefont
  {A.}~\bibnamefont {Bansil}}, \bibinfo {author} {\bibfnamefont
  {F.}~\bibnamefont {Chou}}, \bibinfo {author} {\bibfnamefont {P.~P.}\
  \bibnamefont {Shibayev}}, \bibinfo {author} {\bibfnamefont {H.}~\bibnamefont
  {Lin}}, \bibinfo {author} {\bibfnamefont {S.}~\bibnamefont {Jia}},\ and\
  \bibinfo {author} {\bibfnamefont {M.~Z.}\ \bibnamefont {Hasan}},\ }\bibfield
  {title} {\bibinfo {title} {Discovery of a weyl fermion semimetal and
  topological fermi arcs},\ }\href {https://doi.org/10.1126/science.aaa9297}
  {\bibfield  {journal} {\bibinfo  {journal} {Science}\ }\textbf {\bibinfo
  {volume} {349}},\ \bibinfo {pages} {613} (\bibinfo {year}
  {2015})}\BibitemShut {NoStop}%
\bibitem [{\citenamefont {Yang}\ \emph {et~al.}(2015)\citenamefont {Yang},
  \citenamefont {Liu}, \citenamefont {Sun}, \citenamefont {Peng}, \citenamefont
  {Yang}, \citenamefont {Zhang}, \citenamefont {Zhou}, \citenamefont {Zhang},
  \citenamefont {Guo}, \citenamefont {Rahn}, \citenamefont {Prabhakaran},
  \citenamefont {Hussain}, \citenamefont {Mo}, \citenamefont {Felser},
  \citenamefont {Yan},\ and\ \citenamefont {Chen}}]{Yang2015}%
  \BibitemOpen
  \bibfield  {author} {\bibinfo {author} {\bibfnamefont {L.~X.}\ \bibnamefont
  {Yang}}, \bibinfo {author} {\bibfnamefont {Z.~K.}\ \bibnamefont {Liu}},
  \bibinfo {author} {\bibfnamefont {Y.}~\bibnamefont {Sun}}, \bibinfo {author}
  {\bibfnamefont {H.}~\bibnamefont {Peng}}, \bibinfo {author} {\bibfnamefont
  {H.~F.}\ \bibnamefont {Yang}}, \bibinfo {author} {\bibfnamefont
  {T.}~\bibnamefont {Zhang}}, \bibinfo {author} {\bibfnamefont
  {B.}~\bibnamefont {Zhou}}, \bibinfo {author} {\bibfnamefont {Y.}~\bibnamefont
  {Zhang}}, \bibinfo {author} {\bibfnamefont {Y.~F.}\ \bibnamefont {Guo}},
  \bibinfo {author} {\bibfnamefont {M.}~\bibnamefont {Rahn}}, \bibinfo {author}
  {\bibfnamefont {D.}~\bibnamefont {Prabhakaran}}, \bibinfo {author}
  {\bibfnamefont {Z.}~\bibnamefont {Hussain}}, \bibinfo {author} {\bibfnamefont
  {S.-K.}\ \bibnamefont {Mo}}, \bibinfo {author} {\bibfnamefont
  {C.}~\bibnamefont {Felser}}, \bibinfo {author} {\bibfnamefont
  {B.}~\bibnamefont {Yan}},\ and\ \bibinfo {author} {\bibfnamefont {Y.~L.}\
  \bibnamefont {Chen}},\ }\bibfield  {title} {\bibinfo {title} {Weyl semimetal
  phase in the non-centrosymmetric compound taas},\ }\href
  {https://doi.org/10.1038/nphys3425} {\bibfield  {journal} {\bibinfo
  {journal} {Nature Physics}\ }\textbf {\bibinfo {volume} {11}},\ \bibinfo
  {pages} {728} (\bibinfo {year} {2015})}\BibitemShut {NoStop}%
\bibitem [{\citenamefont {Qi}\ and\ \citenamefont
  {Zhang}(2011)}]{RevModPhys.83.1057}%
  \BibitemOpen
  \bibfield  {author} {\bibinfo {author} {\bibfnamefont {X.-L.}\ \bibnamefont
  {Qi}}\ and\ \bibinfo {author} {\bibfnamefont {S.-C.}\ \bibnamefont {Zhang}},\
  }\bibfield  {title} {\bibinfo {title} {Topological insulators and
  superconductors},\ }\href {https://doi.org/10.1103/RevModPhys.83.1057}
  {\bibfield  {journal} {\bibinfo  {journal} {Rev. Mod. Phys.}\ }\textbf
  {\bibinfo {volume} {83}},\ \bibinfo {pages} {1057} (\bibinfo {year}
  {2011})}\BibitemShut {NoStop}%
\bibitem [{\citenamefont {Hasan}\ and\ \citenamefont
  {Kane}(2010)}]{RevModPhys.82.3045}%
  \BibitemOpen
  \bibfield  {author} {\bibinfo {author} {\bibfnamefont {M.~Z.}\ \bibnamefont
  {Hasan}}\ and\ \bibinfo {author} {\bibfnamefont {C.~L.}\ \bibnamefont
  {Kane}},\ }\bibfield  {title} {\bibinfo {title} {Colloquium: Topological
  insulators},\ }\href {https://doi.org/10.1103/RevModPhys.82.3045} {\bibfield
  {journal} {\bibinfo  {journal} {Rev. Mod. Phys.}\ }\textbf {\bibinfo {volume}
  {82}},\ \bibinfo {pages} {3045} (\bibinfo {year} {2010})}\BibitemShut
  {NoStop}%
\bibitem [{\citenamefont {Wan}\ \emph {et~al.}(2011)\citenamefont {Wan},
  \citenamefont {Turner}, \citenamefont {Vishwanath},\ and\ \citenamefont
  {Savrasov}}]{PhysRevB.83.205101}%
  \BibitemOpen
  \bibfield  {author} {\bibinfo {author} {\bibfnamefont {X.}~\bibnamefont
  {Wan}}, \bibinfo {author} {\bibfnamefont {A.~M.}\ \bibnamefont {Turner}},
  \bibinfo {author} {\bibfnamefont {A.}~\bibnamefont {Vishwanath}},\ and\
  \bibinfo {author} {\bibfnamefont {S.~Y.}\ \bibnamefont {Savrasov}},\
  }\bibfield  {title} {\bibinfo {title} {Topological semimetal and fermi-arc
  surface states in the electronic structure of pyrochlore iridates},\ }\href
  {https://doi.org/10.1103/PhysRevB.83.205101} {\bibfield  {journal} {\bibinfo
  {journal} {Phys. Rev. B}\ }\textbf {\bibinfo {volume} {83}},\ \bibinfo
  {pages} {205101} (\bibinfo {year} {2011})}\BibitemShut {NoStop}%
\bibitem [{\citenamefont {Burkov}\ and\ \citenamefont
  {Balents}(2011)}]{PhysRevLett.107.127205}%
  \BibitemOpen
  \bibfield  {author} {\bibinfo {author} {\bibfnamefont {A.~A.}\ \bibnamefont
  {Burkov}}\ and\ \bibinfo {author} {\bibfnamefont {L.}~\bibnamefont
  {Balents}},\ }\bibfield  {title} {\bibinfo {title} {Weyl semimetal in a
  topological insulator multilayer},\ }\href
  {https://doi.org/10.1103/PhysRevLett.107.127205} {\bibfield  {journal}
  {\bibinfo  {journal} {Phys. Rev. Lett.}\ }\textbf {\bibinfo {volume} {107}},\
  \bibinfo {pages} {127205} (\bibinfo {year} {2011})}\BibitemShut {NoStop}%
\bibitem [{\citenamefont {Nakatsuji}\ \emph {et~al.}(2015)\citenamefont
  {Nakatsuji}, \citenamefont {Kiyohara},\ and\ \citenamefont
  {Higo}}]{Nakatsuji2015}%
  \BibitemOpen
  \bibfield  {author} {\bibinfo {author} {\bibfnamefont {S.}~\bibnamefont
  {Nakatsuji}}, \bibinfo {author} {\bibfnamefont {N.}~\bibnamefont
  {Kiyohara}},\ and\ \bibinfo {author} {\bibfnamefont {T.}~\bibnamefont
  {Higo}},\ }\bibfield  {title} {\bibinfo {title} {Large anomalous hall effect
  in a non-collinear antiferromagnet at room temperature},\ }\href
  {https://doi.org/10.1038/nature15723} {\bibfield  {journal} {\bibinfo
  {journal} {Nature}\ }\textbf {\bibinfo {volume} {527}},\ \bibinfo {pages}
  {212} (\bibinfo {year} {2015})}\BibitemShut {NoStop}%
\bibitem [{\citenamefont {Rout}\ \emph {et~al.}(2019)\citenamefont {Rout},
  \citenamefont {Madduri}, \citenamefont {Manna},\ and\ \citenamefont
  {Nayak}}]{PhysRevB.99.094430}%
  \BibitemOpen
  \bibfield  {author} {\bibinfo {author} {\bibfnamefont {P.~K.}\ \bibnamefont
  {Rout}}, \bibinfo {author} {\bibfnamefont {P.~V.~P.}\ \bibnamefont
  {Madduri}}, \bibinfo {author} {\bibfnamefont {S.~K.}\ \bibnamefont {Manna}},\
  and\ \bibinfo {author} {\bibfnamefont {A.~K.}\ \bibnamefont {Nayak}},\
  }\bibfield  {title} {\bibinfo {title} {Field-induced topological hall effect
  in the noncoplanar triangular antiferromagnetic geometry of
  ${\mathrm{\uppercase{m}n}}_{3}\mathrm{\uppercase{s}n}$},\ }\href
  {https://doi.org/10.1103/PhysRevB.99.094430} {\bibfield  {journal} {\bibinfo
  {journal} {Phys. Rev. B}\ }\textbf {\bibinfo {volume} {99}},\ \bibinfo
  {pages} {094430} (\bibinfo {year} {2019})}\BibitemShut {NoStop}%
\bibitem [{\citenamefont {Kimata}\ \emph {et~al.}(2019)\citenamefont {Kimata},
  \citenamefont {Chen}, \citenamefont {Kondou}, \citenamefont {Sugimoto},
  \citenamefont {Muduli}, \citenamefont {Ikhlas}, \citenamefont {Omori},
  \citenamefont {Tomita}, \citenamefont {MacDonald}, \citenamefont
  {Nakatsuji},\ and\ \citenamefont {Otani}}]{Kimata2019}%
  \BibitemOpen
  \bibfield  {author} {\bibinfo {author} {\bibfnamefont {M.}~\bibnamefont
  {Kimata}}, \bibinfo {author} {\bibfnamefont {H.}~\bibnamefont {Chen}},
  \bibinfo {author} {\bibfnamefont {K.}~\bibnamefont {Kondou}}, \bibinfo
  {author} {\bibfnamefont {S.}~\bibnamefont {Sugimoto}}, \bibinfo {author}
  {\bibfnamefont {P.~K.}\ \bibnamefont {Muduli}}, \bibinfo {author}
  {\bibfnamefont {M.}~\bibnamefont {Ikhlas}}, \bibinfo {author} {\bibfnamefont
  {Y.}~\bibnamefont {Omori}}, \bibinfo {author} {\bibfnamefont
  {T.}~\bibnamefont {Tomita}}, \bibinfo {author} {\bibfnamefont {A.~H.}\
  \bibnamefont {MacDonald}}, \bibinfo {author} {\bibfnamefont {S.}~\bibnamefont
  {Nakatsuji}},\ and\ \bibinfo {author} {\bibfnamefont {Y.}~\bibnamefont
  {Otani}},\ }\bibfield  {title} {\bibinfo {title} {Magnetic and magnetic
  inverse spin hall effects in a non-collinear antiferromagnet},\ }\href
  {https://doi.org/10.1038/s41586-018-0853-0} {\bibfield  {journal} {\bibinfo
  {journal} {Nature}\ }\textbf {\bibinfo {volume} {565}},\ \bibinfo {pages}
  {627} (\bibinfo {year} {2019})}\BibitemShut {NoStop}%
\bibitem [{\citenamefont {Higo}\ \emph {et~al.}(2018)\citenamefont {Higo},
  \citenamefont {Man}, \citenamefont {Gopman}, \citenamefont {Wu},
  \citenamefont {Koretsune}, \citenamefont {van~'t Erve}, \citenamefont
  {Kabanov}, \citenamefont {Rees}, \citenamefont {Li}, \citenamefont {Suzuki},
  \citenamefont {Patankar}, \citenamefont {Ikhlas}, \citenamefont {Chien},
  \citenamefont {Arita}, \citenamefont {Shull}, \citenamefont {Orenstein},\
  and\ \citenamefont {Nakatsuji}}]{Higo2018}%
  \BibitemOpen
  \bibfield  {author} {\bibinfo {author} {\bibfnamefont {T.}~\bibnamefont
  {Higo}}, \bibinfo {author} {\bibfnamefont {H.}~\bibnamefont {Man}}, \bibinfo
  {author} {\bibfnamefont {D.~B.}\ \bibnamefont {Gopman}}, \bibinfo {author}
  {\bibfnamefont {L.}~\bibnamefont {Wu}}, \bibinfo {author} {\bibfnamefont
  {T.}~\bibnamefont {Koretsune}}, \bibinfo {author} {\bibfnamefont {O.~M.~J.}\
  \bibnamefont {van~'t Erve}}, \bibinfo {author} {\bibfnamefont {Y.~P.}\
  \bibnamefont {Kabanov}}, \bibinfo {author} {\bibfnamefont {D.}~\bibnamefont
  {Rees}}, \bibinfo {author} {\bibfnamefont {Y.}~\bibnamefont {Li}}, \bibinfo
  {author} {\bibfnamefont {M.-T.}\ \bibnamefont {Suzuki}}, \bibinfo {author}
  {\bibfnamefont {S.}~\bibnamefont {Patankar}}, \bibinfo {author}
  {\bibfnamefont {M.}~\bibnamefont {Ikhlas}}, \bibinfo {author} {\bibfnamefont
  {C.~L.}\ \bibnamefont {Chien}}, \bibinfo {author} {\bibfnamefont
  {R.}~\bibnamefont {Arita}}, \bibinfo {author} {\bibfnamefont {R.~D.}\
  \bibnamefont {Shull}}, \bibinfo {author} {\bibfnamefont {J.}~\bibnamefont
  {Orenstein}},\ and\ \bibinfo {author} {\bibfnamefont {S.}~\bibnamefont
  {Nakatsuji}},\ }\bibfield  {title} {\bibinfo {title} {Large magneto-optical
  kerr effect and imaging of magnetic octupole domains in an antiferromagnetic
  metal},\ }\href {https://doi.org/10.1038/s41566-017-0086-z} {\bibfield
  {journal} {\bibinfo  {journal} {Nature Photonics}\ }\textbf {\bibinfo
  {volume} {12}},\ \bibinfo {pages} {73} (\bibinfo {year} {2018})}\BibitemShut
  {NoStop}%
\bibitem [{\citenamefont {Brown}\ \emph {et~al.}(1990)\citenamefont {Brown},
  \citenamefont {Nunez}, \citenamefont {Tasset}, \citenamefont {Forsyth},\ and\
  \citenamefont {Radhakrishna}}]{P_J_Brown_1990}%
  \BibitemOpen
  \bibfield  {author} {\bibinfo {author} {\bibfnamefont {P.~J.}\ \bibnamefont
  {Brown}}, \bibinfo {author} {\bibfnamefont {V.}~\bibnamefont {Nunez}},
  \bibinfo {author} {\bibfnamefont {F.}~\bibnamefont {Tasset}}, \bibinfo
  {author} {\bibfnamefont {J.~B.}\ \bibnamefont {Forsyth}},\ and\ \bibinfo
  {author} {\bibfnamefont {P.}~\bibnamefont {Radhakrishna}},\ }\bibfield
  {title} {\bibinfo {title} {Determination of the magnetic structure of
  $\mathrm{Mn}_3\mathrm{Sn}$ using generalized neutron polarization analysis},\
  }\href {https://doi.org/10.1088/0953-8984/2/47/015} {\bibfield  {journal}
  {\bibinfo  {journal} {Journal of Physics: Condensed Matter}\ }\textbf
  {\bibinfo {volume} {2}},\ \bibinfo {pages} {9409} (\bibinfo {year}
  {1990})}\BibitemShut {NoStop}%
\bibitem [{\citenamefont {Tomiyoshi}\ and\ \citenamefont
  {Yamaguchi}(1982{\natexlab{a}})}]{Tomiyoshi_Yamaguchi_1982}%
  \BibitemOpen
  \bibfield  {author} {\bibinfo {author} {\bibfnamefont {S.}~\bibnamefont
  {Tomiyoshi}}\ and\ \bibinfo {author} {\bibfnamefont {Y.}~\bibnamefont
  {Yamaguchi}},\ }\bibfield  {title} {\bibinfo {title} {Magnetic structure and
  weak ferromagnetism of $\mathrm{Mn}_3\mathrm{Sn}$ studied by polarized
  neutron diffraction},\ }\href {https://doi.org/10.1143/jpsj.51.2478}
  {\bibfield  {journal} {\bibinfo  {journal} {Journal of the Physical Society
  of Japan}\ }\textbf {\bibinfo {volume} {51}},\ \bibinfo {pages} {2478–2486}
  (\bibinfo {year} {1982}{\natexlab{a}})}\BibitemShut {NoStop}%
\bibitem [{\citenamefont {Song}\ \emph {et~al.}(2020)\citenamefont {Song},
  \citenamefont {Hao}, \citenamefont {Wang}, \citenamefont {Zhang},
  \citenamefont {Huang}, \citenamefont {Xing},\ and\ \citenamefont
  {Chen}}]{PhysRevB.101.144422}%
  \BibitemOpen
  \bibfield  {author} {\bibinfo {author} {\bibfnamefont {Y.}~\bibnamefont
  {Song}}, \bibinfo {author} {\bibfnamefont {Y.}~\bibnamefont {Hao}}, \bibinfo
  {author} {\bibfnamefont {S.}~\bibnamefont {Wang}}, \bibinfo {author}
  {\bibfnamefont {J.}~\bibnamefont {Zhang}}, \bibinfo {author} {\bibfnamefont
  {Q.}~\bibnamefont {Huang}}, \bibinfo {author} {\bibfnamefont
  {X.}~\bibnamefont {Xing}},\ and\ \bibinfo {author} {\bibfnamefont
  {J.}~\bibnamefont {Chen}},\ }\bibfield  {title} {\bibinfo {title}
  {Complicated magnetic structure and its strong correlation with the anomalous
  hall effect in ${\mathrm{\uppercase{m}n}}_{3}\mathrm{\uppercase{s}n}$},\
  }\href {https://doi.org/10.1103/PhysRevB.101.144422} {\bibfield  {journal}
  {\bibinfo  {journal} {Phys. Rev. B}\ }\textbf {\bibinfo {volume} {101}},\
  \bibinfo {pages} {144422} (\bibinfo {year} {2020})}\BibitemShut {NoStop}%
\bibitem [{\citenamefont {Wang}\ \emph {et~al.}(2023)\citenamefont {Wang},
  \citenamefont {Zhu}, \citenamefont {Yang}, \citenamefont {Meven},
  \citenamefont {Mi}, \citenamefont {Yi}, \citenamefont {Song}, \citenamefont
  {Mueller}, \citenamefont {Schmidt}, \citenamefont {Schmalzl}, \citenamefont
  {Ressouche}, \citenamefont {Xu}, \citenamefont {He}, \citenamefont {Shi},
  \citenamefont {Feng}, \citenamefont {Mokrousov}, \citenamefont {Blügel},
  \citenamefont {Roth},\ and\ \citenamefont {Su}}]{wang2023flat}%
  \BibitemOpen
  \bibfield  {author} {\bibinfo {author} {\bibfnamefont {X.}~\bibnamefont
  {Wang}}, \bibinfo {author} {\bibfnamefont {F.}~\bibnamefont {Zhu}}, \bibinfo
  {author} {\bibfnamefont {X.}~\bibnamefont {Yang}}, \bibinfo {author}
  {\bibfnamefont {M.}~\bibnamefont {Meven}}, \bibinfo {author} {\bibfnamefont
  {X.}~\bibnamefont {Mi}}, \bibinfo {author} {\bibfnamefont {C.}~\bibnamefont
  {Yi}}, \bibinfo {author} {\bibfnamefont {J.}~\bibnamefont {Song}}, \bibinfo
  {author} {\bibfnamefont {T.}~\bibnamefont {Mueller}}, \bibinfo {author}
  {\bibfnamefont {W.}~\bibnamefont {Schmidt}}, \bibinfo {author} {\bibfnamefont
  {K.}~\bibnamefont {Schmalzl}}, \bibinfo {author} {\bibfnamefont
  {E.}~\bibnamefont {Ressouche}}, \bibinfo {author} {\bibfnamefont
  {J.}~\bibnamefont {Xu}}, \bibinfo {author} {\bibfnamefont {M.}~\bibnamefont
  {He}}, \bibinfo {author} {\bibfnamefont {Y.}~\bibnamefont {Shi}}, \bibinfo
  {author} {\bibfnamefont {W.}~\bibnamefont {Feng}}, \bibinfo {author}
  {\bibfnamefont {Y.}~\bibnamefont {Mokrousov}}, \bibinfo {author}
  {\bibfnamefont {S.}~\bibnamefont {Blügel}}, \bibinfo {author} {\bibfnamefont
  {G.}~\bibnamefont {Roth}},\ and\ \bibinfo {author} {\bibfnamefont
  {Y.}~\bibnamefont {Su}},\ }\bibfield  {title} {\bibinfo {title} {Flat
  band-engineered spin-density wave and the emergent multi-$k$ magnetic state
  in the topological kagome metal \uppercase{M}n$_{3}$\uppercase{S}n},\
  }\href@noop {} {\  (\bibinfo {year} {2023})},\ \Eprint
  {https://arxiv.org/abs/2306.04312} {arXiv:2306.04312 [cond-mat.str-el]}
  \BibitemShut {NoStop}%
\bibitem [{\citenamefont {Park}\ \emph {et~al.}(2018)\citenamefont {Park},
  \citenamefont {Oh}, \citenamefont {Uhl{\'i}{\v{r}}ov{\'a}}, \citenamefont
  {Jackson}, \citenamefont {De{\'a}k}, \citenamefont {Szunyogh}, \citenamefont
  {Lee}, \citenamefont {Cho}, \citenamefont {Kim}, \citenamefont {Walker},
  \citenamefont {Adroja}, \citenamefont {Sechovsk{\'y}},\ and\ \citenamefont
  {Park}}]{Park2018}%
  \BibitemOpen
  \bibfield  {author} {\bibinfo {author} {\bibfnamefont {P.}~\bibnamefont
  {Park}}, \bibinfo {author} {\bibfnamefont {J.}~\bibnamefont {Oh}}, \bibinfo
  {author} {\bibfnamefont {K.}~\bibnamefont {Uhl{\'i}{\v{r}}ov{\'a}}}, \bibinfo
  {author} {\bibfnamefont {J.}~\bibnamefont {Jackson}}, \bibinfo {author}
  {\bibfnamefont {A.}~\bibnamefont {De{\'a}k}}, \bibinfo {author}
  {\bibfnamefont {L.}~\bibnamefont {Szunyogh}}, \bibinfo {author}
  {\bibfnamefont {K.~H.}\ \bibnamefont {Lee}}, \bibinfo {author} {\bibfnamefont
  {H.}~\bibnamefont {Cho}}, \bibinfo {author} {\bibfnamefont {H.-L.}\
  \bibnamefont {Kim}}, \bibinfo {author} {\bibfnamefont {H.~C.}\ \bibnamefont
  {Walker}}, \bibinfo {author} {\bibfnamefont {D.}~\bibnamefont {Adroja}},
  \bibinfo {author} {\bibfnamefont {V.}~\bibnamefont {Sechovsk{\'y}}},\ and\
  \bibinfo {author} {\bibfnamefont {J.-G.}\ \bibnamefont {Park}},\ }\bibfield
  {title} {\bibinfo {title} {Magnetic excitations in non-collinear
  antiferromagnetic weyl semimetal $\mathrm{Mn}_3\mathrm{Sn}$},\ }\href
  {https://doi.org/10.1038/s41535-018-0137-9} {\bibfield  {journal} {\bibinfo
  {journal} {npj Quantum Materials}\ }\textbf {\bibinfo {volume} {3}},\
  \bibinfo {pages} {63} (\bibinfo {year} {2018})}\BibitemShut {NoStop}%
\bibitem [{\citenamefont {Zhang}\ \emph {et~al.}(2017)\citenamefont {Zhang},
  \citenamefont {Sun}, \citenamefont {Yang}, \citenamefont
  {\ifmmode~\check{Z}\else \v{Z}\fi{}elezn\'y}, \citenamefont {Parkin},
  \citenamefont {Felser},\ and\ \citenamefont {Yan}}]{PhysRevB.95.075128}%
  \BibitemOpen
  \bibfield  {author} {\bibinfo {author} {\bibfnamefont {Y.}~\bibnamefont
  {Zhang}}, \bibinfo {author} {\bibfnamefont {Y.}~\bibnamefont {Sun}}, \bibinfo
  {author} {\bibfnamefont {H.}~\bibnamefont {Yang}}, \bibinfo {author}
  {\bibfnamefont {J.}~\bibnamefont {\ifmmode~\check{Z}\else
  \v{Z}\fi{}elezn\'y}}, \bibinfo {author} {\bibfnamefont {S.~P.~P.}\
  \bibnamefont {Parkin}}, \bibinfo {author} {\bibfnamefont {C.}~\bibnamefont
  {Felser}},\ and\ \bibinfo {author} {\bibfnamefont {B.}~\bibnamefont {Yan}},\
  }\bibfield  {title} {\bibinfo {title} {Strong anisotropic anomalous hall
  effect and spin hall effect in the chiral antiferromagnetic compounds
  $\mathrm{Mn}_3\mathrm{Z}$ ($\mathrm{Z=Ge, Sn, Ga, Ir, Rh}$ and
  $\mathrm{Pt}$)},\ }\href {https://doi.org/10.1103/PhysRevB.95.075128}
  {\bibfield  {journal} {\bibinfo  {journal} {Phys. Rev. B}\ }\textbf {\bibinfo
  {volume} {95}},\ \bibinfo {pages} {075128} (\bibinfo {year}
  {2017})}\BibitemShut {NoStop}%
\bibitem [{\citenamefont {Kübler}\ and\ \citenamefont
  {Felser}(2014)}]{Kübler_2014}%
  \BibitemOpen
  \bibfield  {author} {\bibinfo {author} {\bibfnamefont {J.}~\bibnamefont
  {Kübler}}\ and\ \bibinfo {author} {\bibfnamefont {C.}~\bibnamefont
  {Felser}},\ }\bibfield  {title} {\bibinfo {title} {Non-collinear
  antiferromagnets and the anomalous hall effect},\ }\href
  {https://doi.org/10.1209/0295-5075/108/67001} {\bibfield  {journal} {\bibinfo
   {journal} {Europhysics Letters}\ }\textbf {\bibinfo {volume} {108}},\
  \bibinfo {pages} {67001} (\bibinfo {year} {2014})}\BibitemShut {NoStop}%
\bibitem [{\citenamefont {Pradhan}\ \emph {et~al.}(2023)\citenamefont
  {Pradhan}, \citenamefont {Samanta}, \citenamefont {Saha},\ and\ \citenamefont
  {Nandy}}]{Pradhan2023}%
  \BibitemOpen
  \bibfield  {author} {\bibinfo {author} {\bibfnamefont {S.}~\bibnamefont
  {Pradhan}}, \bibinfo {author} {\bibfnamefont {K.}~\bibnamefont {Samanta}},
  \bibinfo {author} {\bibfnamefont {K.}~\bibnamefont {Saha}},\ and\ \bibinfo
  {author} {\bibfnamefont {A.~K.}\ \bibnamefont {Nandy}},\ }\bibfield  {title}
  {\bibinfo {title} {Vector-chirality driven topological phase transitions in
  noncollinear antiferromagnets and its impact on anomalous hall effect},\
  }\href {https://doi.org/10.1038/s42005-023-01385-9} {\bibfield  {journal}
  {\bibinfo  {journal} {Communications Physics}\ }\textbf {\bibinfo {volume}
  {6}},\ \bibinfo {pages} {272} (\bibinfo {year} {2023})}\BibitemShut {NoStop}%
\bibitem [{\citenamefont {Wang}\ \emph {et~al.}(2019)\citenamefont {Wang},
  \citenamefont {Li}, \citenamefont {Zhang}, \citenamefont {Hou}, \citenamefont
  {Zhao}, \citenamefont {Li}, \citenamefont {Rahman}, \citenamefont {Xu},
  \citenamefont {Gong}, \citenamefont {Chi}, \citenamefont {Dai}, \citenamefont
  {Wang}, \citenamefont {Qiao},\ and\ \citenamefont
  {Zhang}}]{PhysRevB.100.014407}%
  \BibitemOpen
  \bibfield  {author} {\bibinfo {author} {\bibfnamefont {X.}~\bibnamefont
  {Wang}}, \bibinfo {author} {\bibfnamefont {Z.}~\bibnamefont {Li}}, \bibinfo
  {author} {\bibfnamefont {M.}~\bibnamefont {Zhang}}, \bibinfo {author}
  {\bibfnamefont {T.}~\bibnamefont {Hou}}, \bibinfo {author} {\bibfnamefont
  {J.}~\bibnamefont {Zhao}}, \bibinfo {author} {\bibfnamefont {L.}~\bibnamefont
  {Li}}, \bibinfo {author} {\bibfnamefont {A.}~\bibnamefont {Rahman}}, \bibinfo
  {author} {\bibfnamefont {Z.}~\bibnamefont {Xu}}, \bibinfo {author}
  {\bibfnamefont {J.}~\bibnamefont {Gong}}, \bibinfo {author} {\bibfnamefont
  {Z.}~\bibnamefont {Chi}}, \bibinfo {author} {\bibfnamefont {R.}~\bibnamefont
  {Dai}}, \bibinfo {author} {\bibfnamefont {Z.}~\bibnamefont {Wang}}, \bibinfo
  {author} {\bibfnamefont {Z.}~\bibnamefont {Qiao}},\ and\ \bibinfo {author}
  {\bibfnamefont {Z.}~\bibnamefont {Zhang}},\ }\bibfield  {title} {\bibinfo
  {title} {Pressure-induced modification of the anomalous hall effect in
  layered $\mathrm{Fe}_{3}\mathrm{GeTe}_2$},\ }\href
  {https://doi.org/10.1103/PhysRevB.100.014407} {\bibfield  {journal} {\bibinfo
   {journal} {Phys. Rev. B}\ }\textbf {\bibinfo {volume} {100}},\ \bibinfo
  {pages} {014407} (\bibinfo {year} {2019})}\BibitemShut {NoStop}%
\bibitem [{\citenamefont {Sun}\ \emph {et~al.}(2021)\citenamefont {Sun},
  \citenamefont {Peng}, \citenamefont {Cui}, \citenamefont {Zhu}, \citenamefont
  {Zhuo}, \citenamefont {Wang},\ and\ \citenamefont
  {Chen}}]{PhysRevB.103.085116}%
  \BibitemOpen
  \bibfield  {author} {\bibinfo {author} {\bibfnamefont {Z.~L.}\ \bibnamefont
  {Sun}}, \bibinfo {author} {\bibfnamefont {K.~L.}\ \bibnamefont {Peng}},
  \bibinfo {author} {\bibfnamefont {J.~H.}\ \bibnamefont {Cui}}, \bibinfo
  {author} {\bibfnamefont {C.~S.}\ \bibnamefont {Zhu}}, \bibinfo {author}
  {\bibfnamefont {W.~Z.}\ \bibnamefont {Zhuo}}, \bibinfo {author}
  {\bibfnamefont {Z.~Y.}\ \bibnamefont {Wang}},\ and\ \bibinfo {author}
  {\bibfnamefont {X.~H.}\ \bibnamefont {Chen}},\ }\bibfield  {title} {\bibinfo
  {title} {Pressure-controlled anomalous hall conductivity in the half-heusler
  antiferromagnet $\mathrm{GdPtBi}$},\ }\href
  {https://doi.org/10.1103/PhysRevB.103.085116} {\bibfield  {journal} {\bibinfo
   {journal} {Phys. Rev. B}\ }\textbf {\bibinfo {volume} {103}},\ \bibinfo
  {pages} {085116} (\bibinfo {year} {2021})}\BibitemShut {NoStop}%
\bibitem [{\citenamefont {Rai}\ \emph {et~al.}(2023)\citenamefont {Rai},
  \citenamefont {Stunault}, \citenamefont {Schmidt}, \citenamefont {Jana},
  \citenamefont {Per\ss{}on}, \citenamefont {Soh}, \citenamefont {Br\"uckel},\
  and\ \citenamefont {Nandi}}]{PhysRevB.107.184413}%
  \BibitemOpen
  \bibfield  {author} {\bibinfo {author} {\bibfnamefont {V.}~\bibnamefont
  {Rai}}, \bibinfo {author} {\bibfnamefont {A.}~\bibnamefont {Stunault}},
  \bibinfo {author} {\bibfnamefont {W.}~\bibnamefont {Schmidt}}, \bibinfo
  {author} {\bibfnamefont {S.}~\bibnamefont {Jana}}, \bibinfo {author}
  {\bibfnamefont {J.}~\bibnamefont {Per\ss{}on}}, \bibinfo {author}
  {\bibfnamefont {J.-R.}\ \bibnamefont {Soh}}, \bibinfo {author} {\bibfnamefont
  {T.}~\bibnamefont {Br\"uckel}},\ and\ \bibinfo {author} {\bibfnamefont
  {S.}~\bibnamefont {Nandi}},\ }\bibfield  {title} {\bibinfo {title} {Anomalous
  hall effect and magnetic structure of the topological semimetal
  ($\mathrm{Mn}_{0.78}\mathrm{Fe}_{0.22})_{3}\mathrm{Ge}$},\ }\href
  {https://doi.org/10.1103/PhysRevB.107.184413} {\bibfield  {journal} {\bibinfo
   {journal} {Phys. Rev. B}\ }\textbf {\bibinfo {volume} {107}},\ \bibinfo
  {pages} {184413} (\bibinfo {year} {2023})}\BibitemShut {NoStop}%
\bibitem [{\citenamefont {Shen}\ \emph {et~al.}(2020)\citenamefont {Shen},
  \citenamefont {Yao}, \citenamefont {Zeng}, \citenamefont {Sun}, \citenamefont
  {Xi}, \citenamefont {Wu}, \citenamefont {Wang}, \citenamefont {Shen},
  \citenamefont {Liu},\ and\ \citenamefont {Liu}}]{PhysRevLett.125.086602}%
  \BibitemOpen
  \bibfield  {author} {\bibinfo {author} {\bibfnamefont {J.}~\bibnamefont
  {Shen}}, \bibinfo {author} {\bibfnamefont {Q.}~\bibnamefont {Yao}}, \bibinfo
  {author} {\bibfnamefont {Q.}~\bibnamefont {Zeng}}, \bibinfo {author}
  {\bibfnamefont {H.}~\bibnamefont {Sun}}, \bibinfo {author} {\bibfnamefont
  {X.}~\bibnamefont {Xi}}, \bibinfo {author} {\bibfnamefont {G.}~\bibnamefont
  {Wu}}, \bibinfo {author} {\bibfnamefont {W.}~\bibnamefont {Wang}}, \bibinfo
  {author} {\bibfnamefont {B.}~\bibnamefont {Shen}}, \bibinfo {author}
  {\bibfnamefont {Q.}~\bibnamefont {Liu}},\ and\ \bibinfo {author}
  {\bibfnamefont {E.}~\bibnamefont {Liu}},\ }\bibfield  {title} {\bibinfo
  {title} {Local disorder-induced elevation of intrinsic anomalous hall
  conductance in an electron-doped magnetic weyl semimetal},\ }\href
  {https://doi.org/10.1103/PhysRevLett.125.086602} {\bibfield  {journal}
  {\bibinfo  {journal} {Phys. Rev. Lett.}\ }\textbf {\bibinfo {volume} {125}},\
  \bibinfo {pages} {086602} (\bibinfo {year} {2020})}\BibitemShut {NoStop}%
\bibitem [{\citenamefont {Shen}\ \emph {et~al.}(2022)\citenamefont {Shen},
  \citenamefont {Zhang}, \citenamefont {Liang}, \citenamefont {Wang},
  \citenamefont {Zeng}, \citenamefont {Wang}, \citenamefont {Wei},
  \citenamefont {Liu},\ and\ \citenamefont {Xu}}]{10.1063/5.0095950}%
  \BibitemOpen
  \bibfield  {author} {\bibinfo {author} {\bibfnamefont {J.}~\bibnamefont
  {Shen}}, \bibinfo {author} {\bibfnamefont {S.}~\bibnamefont {Zhang}},
  \bibinfo {author} {\bibfnamefont {T.}~\bibnamefont {Liang}}, \bibinfo
  {author} {\bibfnamefont {J.}~\bibnamefont {Wang}}, \bibinfo {author}
  {\bibfnamefont {Q.}~\bibnamefont {Zeng}}, \bibinfo {author} {\bibfnamefont
  {Y.}~\bibnamefont {Wang}}, \bibinfo {author} {\bibfnamefont {H.}~\bibnamefont
  {Wei}}, \bibinfo {author} {\bibfnamefont {E.}~\bibnamefont {Liu}},\ and\
  \bibinfo {author} {\bibfnamefont {X.}~\bibnamefont {Xu}},\ }\bibfield
  {title} {\bibinfo {title} {{Intrinsically enhanced anomalous Hall
  conductivity and Hall angle in Sb-doped magnetic Weyl semimetal
  $\mathrm{Co}_3\mathrm{Sn}_2\mathrm{S}_2$}},\ }\href
  {https://doi.org/10.1063/5.0095950} {\bibfield  {journal} {\bibinfo
  {journal} {APL Materials}\ }\textbf {\bibinfo {volume} {10}} (\bibinfo {year}
  {2022})}\BibitemShut {NoStop}%
\bibitem [{\citenamefont {Liu}\ \emph {et~al.}(2018)\citenamefont {Liu},
  \citenamefont {Teng},\ and\ \citenamefont {Li}}]{Liu2018}%
  \BibitemOpen
  \bibfield  {author} {\bibinfo {author} {\bibfnamefont {N.}~\bibnamefont
  {Liu}}, \bibinfo {author} {\bibfnamefont {J.}~\bibnamefont {Teng}},\ and\
  \bibinfo {author} {\bibfnamefont {Y.}~\bibnamefont {Li}},\ }\bibfield
  {title} {\bibinfo {title} {Two-component anomalous hall effect in a
  magnetically doped topological insulator},\ }\href
  {https://doi.org/10.1038/s41467-018-03684-0} {\bibfield  {journal} {\bibinfo
  {journal} {Nature Communications}\ }\textbf {\bibinfo {volume} {9}},\
  \bibinfo {pages} {1282} (\bibinfo {year} {2018})}\BibitemShut {NoStop}%
\bibitem [{\citenamefont {Guguchia}\ \emph {et~al.}(2020)\citenamefont
  {Guguchia}, \citenamefont {Verezhak}, \citenamefont {Gawryluk}, \citenamefont
  {Tsirkin}, \citenamefont {Yin}, \citenamefont {Belopolski}, \citenamefont
  {Zhou}, \citenamefont {Simutis}, \citenamefont {Zhang}, \citenamefont
  {Cochran}, \citenamefont {Chang}, \citenamefont {Pomjakushina}, \citenamefont
  {Keller}, \citenamefont {Skrzeczkowska}, \citenamefont {Wang}, \citenamefont
  {Lei}, \citenamefont {Khasanov}, \citenamefont {Amato}, \citenamefont {Jia},
  \citenamefont {Neupert}, \citenamefont {Luetkens},\ and\ \citenamefont
  {Hasan}}]{Guguchia2020}%
  \BibitemOpen
  \bibfield  {author} {\bibinfo {author} {\bibfnamefont {Z.}~\bibnamefont
  {Guguchia}}, \bibinfo {author} {\bibfnamefont {J.~A.~T.}\ \bibnamefont
  {Verezhak}}, \bibinfo {author} {\bibfnamefont {D.~J.}\ \bibnamefont
  {Gawryluk}}, \bibinfo {author} {\bibfnamefont {S.~S.}\ \bibnamefont
  {Tsirkin}}, \bibinfo {author} {\bibfnamefont {J.-X.}\ \bibnamefont {Yin}},
  \bibinfo {author} {\bibfnamefont {I.}~\bibnamefont {Belopolski}}, \bibinfo
  {author} {\bibfnamefont {H.}~\bibnamefont {Zhou}}, \bibinfo {author}
  {\bibfnamefont {G.}~\bibnamefont {Simutis}}, \bibinfo {author} {\bibfnamefont
  {S.-S.}\ \bibnamefont {Zhang}}, \bibinfo {author} {\bibfnamefont {T.~A.}\
  \bibnamefont {Cochran}}, \bibinfo {author} {\bibfnamefont {G.}~\bibnamefont
  {Chang}}, \bibinfo {author} {\bibfnamefont {E.}~\bibnamefont {Pomjakushina}},
  \bibinfo {author} {\bibfnamefont {L.}~\bibnamefont {Keller}}, \bibinfo
  {author} {\bibfnamefont {Z.}~\bibnamefont {Skrzeczkowska}}, \bibinfo {author}
  {\bibfnamefont {Q.}~\bibnamefont {Wang}}, \bibinfo {author} {\bibfnamefont
  {H.~C.}\ \bibnamefont {Lei}}, \bibinfo {author} {\bibfnamefont
  {R.}~\bibnamefont {Khasanov}}, \bibinfo {author} {\bibfnamefont
  {A.}~\bibnamefont {Amato}}, \bibinfo {author} {\bibfnamefont
  {S.}~\bibnamefont {Jia}}, \bibinfo {author} {\bibfnamefont {T.}~\bibnamefont
  {Neupert}}, \bibinfo {author} {\bibfnamefont {H.}~\bibnamefont {Luetkens}},\
  and\ \bibinfo {author} {\bibfnamefont {M.~Z.}\ \bibnamefont {Hasan}},\
  }\bibfield  {title} {\bibinfo {title} {Tunable anomalous hall conductivity
  through volume-wise magnetic competition in a topological kagome magnet},\
  }\href {https://doi.org/10.1038/s41467-020-14325-w} {\bibfield  {journal}
  {\bibinfo  {journal} {Nature Communications}\ }\textbf {\bibinfo {volume}
  {11}},\ \bibinfo {pages} {559} (\bibinfo {year} {2020})}\BibitemShut
  {NoStop}%
\bibitem [{\citenamefont {Kren}\ \emph
  {et~al.}(1975{\natexlab{a}})\citenamefont {Kren}, \citenamefont {Paitz},
  \citenamefont {Zimmer},\ and\ \citenamefont {Zsoldos}}]{KPZZ1975}%
  \BibitemOpen
  \bibfield  {author} {\bibinfo {author} {\bibfnamefont {E.}~\bibnamefont
  {Kren}}, \bibinfo {author} {\bibfnamefont {J.}~\bibnamefont {Paitz}},
  \bibinfo {author} {\bibfnamefont {G.}~\bibnamefont {Zimmer}},\ and\ \bibinfo
  {author} {\bibfnamefont {E.}~\bibnamefont {Zsoldos}},\ }\bibfield  {title}
  {\bibinfo {title} {Study of the magnetic phase transformation in the
  $\mathrm{Mn}_3\mathrm{Sn}$ phase},\ }\href
  {https://doi.org/10.1016/0378-4363(75)90066-2} {\bibfield  {journal}
  {\bibinfo  {journal} {Physica B+C}\ }\textbf {\bibinfo {volume} {80}},\
  \bibinfo {pages} {226–230} (\bibinfo {year}
  {1975}{\natexlab{a}})}\BibitemShut {NoStop}%
\bibitem [{\citenamefont {Cao}\ \emph {et~al.}(2023)\citenamefont {Cao},
  \citenamefont {Xu}, \citenamefont {Gao}, \citenamefont {Wang}, \citenamefont
  {Zhang}, \citenamefont {Zhang}, \citenamefont {Guo},\ and\ \citenamefont
  {Chen}}]{cao2023optical}%
  \BibitemOpen
  \bibfield  {author} {\bibinfo {author} {\bibfnamefont {L.~Y.}\ \bibnamefont
  {Cao}}, \bibinfo {author} {\bibfnamefont {Z.~A.}\ \bibnamefont {Xu}},
  \bibinfo {author} {\bibfnamefont {B.~X.}\ \bibnamefont {Gao}}, \bibinfo
  {author} {\bibfnamefont {L.}~\bibnamefont {Wang}}, \bibinfo {author}
  {\bibfnamefont {X.~T.}\ \bibnamefont {Zhang}}, \bibinfo {author}
  {\bibfnamefont {X.~Y.}\ \bibnamefont {Zhang}}, \bibinfo {author}
  {\bibfnamefont {Y.~F.}\ \bibnamefont {Guo}},\ and\ \bibinfo {author}
  {\bibfnamefont {R.~Y.}\ \bibnamefont {Chen}},\ }\bibfield  {title} {\bibinfo
  {title} {Optical study of three-dimensional weyl semimetal
  \uppercase{M}n$_3$\uppercase{S}n},\ }\href@noop {} {\  (\bibinfo {year}
  {2023})},\ \Eprint {https://arxiv.org/abs/2306.12180} {arXiv:2306.12180
  [cond-mat.str-el]} \BibitemShut {NoStop}%
\bibitem [{\citenamefont {Ikhlas}\ \emph {et~al.}()\citenamefont {Ikhlas},
  \citenamefont {Tomita},\ and\ \citenamefont
  {Nakatsuji}}]{doi:10.7566/JPSCP.30.011177}%
  \BibitemOpen
  \bibfield  {author} {\bibinfo {author} {\bibfnamefont {M.}~\bibnamefont
  {Ikhlas}}, \bibinfo {author} {\bibfnamefont {T.}~\bibnamefont {Tomita}},\
  and\ \bibinfo {author} {\bibfnamefont {S.}~\bibnamefont {Nakatsuji}},\
  }\bibinfo {title} {Sample quality dependence of the magnetic properties in
  non-collinear antiferromagnet \uppercase{M}n3\uppercase{S}n},\ in\ \href
  {https://doi.org/10.7566/JPSCP.30.011177} {\emph {\bibinfo {booktitle}
  {Proceedings of the International Conference on Strongly Correlated Electron
  Systems (SCES2019)}}},\ \Eprint
  {https://arxiv.org/abs/https://journals.jps.jp/doi/pdf/10.7566/JPSCP.30.011177}
  {https://journals.jps.jp/doi/pdf/10.7566/JPSCP.30.011177} \BibitemShut
  {NoStop}%
\bibitem [{\citenamefont {Singh}\ \emph {et~al.}(2020)\citenamefont {Singh},
  \citenamefont {Singh}, \citenamefont {Pradhan}, \citenamefont {Srihari},
  \citenamefont {Poswal}, \citenamefont {Nath}, \citenamefont {Nandy},\ and\
  \citenamefont {Nayak}}]{Nayak2020}%
  \BibitemOpen
  \bibfield  {author} {\bibinfo {author} {\bibfnamefont {C.}~\bibnamefont
  {Singh}}, \bibinfo {author} {\bibfnamefont {V.}~\bibnamefont {Singh}},
  \bibinfo {author} {\bibfnamefont {G.}~\bibnamefont {Pradhan}}, \bibinfo
  {author} {\bibfnamefont {V.}~\bibnamefont {Srihari}}, \bibinfo {author}
  {\bibfnamefont {H.~K.}\ \bibnamefont {Poswal}}, \bibinfo {author}
  {\bibfnamefont {R.}~\bibnamefont {Nath}}, \bibinfo {author} {\bibfnamefont
  {A.~K.}\ \bibnamefont {Nandy}},\ and\ \bibinfo {author} {\bibfnamefont
  {A.~K.}\ \bibnamefont {Nayak}},\ }\bibfield  {title} {\bibinfo {title}
  {Pressure controlled trimerization for switching of anomalous hall effect in
  triangular antiferromagnet $\mathrm{Mn}_3\mathrm{Sn}$},\ }\href
  {https://doi.org/10.1103/PhysRevResearch.2.043366} {\bibfield  {journal}
  {\bibinfo  {journal} {Phys. Rev. Res.}\ }\textbf {\bibinfo {volume} {2}},\
  \bibinfo {pages} {043366} (\bibinfo {year} {2020})}\BibitemShut {NoStop}%
\bibitem [{\citenamefont {Hu}\ \emph {et~al.}(2022)\citenamefont {Hu},
  \citenamefont {Wu}, \citenamefont {Ortiz}, \citenamefont {Ju}, \citenamefont
  {Han}, \citenamefont {Ma}, \citenamefont {Plumb}, \citenamefont {Radovic},
  \citenamefont {Thomale}, \citenamefont {Wilson}, \citenamefont {Schnyder},\
  and\ \citenamefont {Shi}}]{Hu2022}%
  \BibitemOpen
  \bibfield  {author} {\bibinfo {author} {\bibfnamefont {Y.}~\bibnamefont
  {Hu}}, \bibinfo {author} {\bibfnamefont {X.}~\bibnamefont {Wu}}, \bibinfo
  {author} {\bibfnamefont {B.~R.}\ \bibnamefont {Ortiz}}, \bibinfo {author}
  {\bibfnamefont {S.}~\bibnamefont {Ju}}, \bibinfo {author} {\bibfnamefont
  {X.}~\bibnamefont {Han}}, \bibinfo {author} {\bibfnamefont {J.}~\bibnamefont
  {Ma}}, \bibinfo {author} {\bibfnamefont {N.~C.}\ \bibnamefont {Plumb}},
  \bibinfo {author} {\bibfnamefont {M.}~\bibnamefont {Radovic}}, \bibinfo
  {author} {\bibfnamefont {R.}~\bibnamefont {Thomale}}, \bibinfo {author}
  {\bibfnamefont {S.~D.}\ \bibnamefont {Wilson}}, \bibinfo {author}
  {\bibfnamefont {A.~P.}\ \bibnamefont {Schnyder}},\ and\ \bibinfo {author}
  {\bibfnamefont {M.}~\bibnamefont {Shi}},\ }\bibfield  {title} {\bibinfo
  {title} {Rich nature of van hove singularities in kagome superconductor
  csv3sb5},\ }\href {https://doi.org/10.1038/s41467-022-29828-x} {\bibfield
  {journal} {\bibinfo  {journal} {Nature Communications}\ }\textbf {\bibinfo
  {volume} {13}},\ \bibinfo {pages} {2220} (\bibinfo {year}
  {2022})}\BibitemShut {NoStop}%
\bibitem [{\citenamefont {Owerre}(2018)}]{PhysRevB.97.094412}%
  \BibitemOpen
  \bibfield  {author} {\bibinfo {author} {\bibfnamefont {S.~A.}\ \bibnamefont
  {Owerre}},\ }\bibfield  {title} {\bibinfo {title} {Weyl magnons in
  noncoplanar stacked kagome antiferromagnets},\ }\href
  {https://doi.org/10.1103/PhysRevB.97.094412} {\bibfield  {journal} {\bibinfo
  {journal} {Phys. Rev. B}\ }\textbf {\bibinfo {volume} {97}},\ \bibinfo
  {pages} {094412} (\bibinfo {year} {2018})}\BibitemShut {NoStop}%
\bibitem [{\citenamefont {Kang}\ \emph {et~al.}(2020)\citenamefont {Kang},
  \citenamefont {Fang}, \citenamefont {Ye}, \citenamefont {Po}, \citenamefont
  {Denlinger}, \citenamefont {Jozwiak}, \citenamefont {Bostwick}, \citenamefont
  {Rotenberg}, \citenamefont {Kaxiras}, \citenamefont {Checkelsky},\ and\
  \citenamefont {Comin}}]{Kang2020}%
  \BibitemOpen
  \bibfield  {author} {\bibinfo {author} {\bibfnamefont {M.}~\bibnamefont
  {Kang}}, \bibinfo {author} {\bibfnamefont {S.}~\bibnamefont {Fang}}, \bibinfo
  {author} {\bibfnamefont {L.}~\bibnamefont {Ye}}, \bibinfo {author}
  {\bibfnamefont {H.~C.}\ \bibnamefont {Po}}, \bibinfo {author} {\bibfnamefont
  {J.}~\bibnamefont {Denlinger}}, \bibinfo {author} {\bibfnamefont
  {C.}~\bibnamefont {Jozwiak}}, \bibinfo {author} {\bibfnamefont
  {A.}~\bibnamefont {Bostwick}}, \bibinfo {author} {\bibfnamefont
  {E.}~\bibnamefont {Rotenberg}}, \bibinfo {author} {\bibfnamefont
  {E.}~\bibnamefont {Kaxiras}}, \bibinfo {author} {\bibfnamefont {J.~G.}\
  \bibnamefont {Checkelsky}},\ and\ \bibinfo {author} {\bibfnamefont
  {R.}~\bibnamefont {Comin}},\ }\bibfield  {title} {\bibinfo {title}
  {Topological flat bands in frustrated kagome lattice
  \uppercase{C}o\uppercase{S}n},\ }\href
  {https://doi.org/10.1038/s41467-020-17465-1} {\bibfield  {journal} {\bibinfo
  {journal} {Nature Communications}\ }\textbf {\bibinfo {volume} {11}},\
  \bibinfo {pages} {4004} (\bibinfo {year} {2020})}\BibitemShut {NoStop}%
\bibitem [{\citenamefont {Choudhury}\ and\ \citenamefont
  {Vorontsov}(2021)}]{PhysRevB.103.104501}%
  \BibitemOpen
  \bibfield  {author} {\bibinfo {author} {\bibfnamefont {S.~S.}\ \bibnamefont
  {Choudhury}}\ and\ \bibinfo {author} {\bibfnamefont {A.~B.}\ \bibnamefont
  {Vorontsov}},\ }\bibfield  {title} {\bibinfo {title} {Thermal transport in
  superconductors with coexisting spin density wave order},\ }\href
  {https://doi.org/10.1103/PhysRevB.103.104501} {\bibfield  {journal} {\bibinfo
   {journal} {Phys. Rev. B}\ }\textbf {\bibinfo {volume} {103}},\ \bibinfo
  {pages} {104501} (\bibinfo {year} {2021})}\BibitemShut {NoStop}%
\bibitem [{\citenamefont {Vorontsov}\ \emph {et~al.}(2010)\citenamefont
  {Vorontsov}, \citenamefont {Vavilov},\ and\ \citenamefont
  {Chubukov}}]{PhysRevB.81.174538}%
  \BibitemOpen
  \bibfield  {author} {\bibinfo {author} {\bibfnamefont {A.~B.}\ \bibnamefont
  {Vorontsov}}, \bibinfo {author} {\bibfnamefont {M.~G.}\ \bibnamefont
  {Vavilov}},\ and\ \bibinfo {author} {\bibfnamefont {A.~V.}\ \bibnamefont
  {Chubukov}},\ }\bibfield  {title} {\bibinfo {title} {Superconductivity and
  spin-density waves in multiband metals},\ }\href
  {https://doi.org/10.1103/PhysRevB.81.174538} {\bibfield  {journal} {\bibinfo
  {journal} {Phys. Rev. B}\ }\textbf {\bibinfo {volume} {81}},\ \bibinfo
  {pages} {174538} (\bibinfo {year} {2010})}\BibitemShut {NoStop}%
\bibitem [{\citenamefont {Tam}\ \emph {et~al.}(2022)\citenamefont {Tam},
  \citenamefont {Zhu}, \citenamefont {Ayres}, \citenamefont {Kummer},
  \citenamefont {Yakhou-Harris}, \citenamefont {Cooper}, \citenamefont
  {Carrington},\ and\ \citenamefont {Hayden}}]{Tam2022}%
  \BibitemOpen
  \bibfield  {author} {\bibinfo {author} {\bibfnamefont {C.~C.}\ \bibnamefont
  {Tam}}, \bibinfo {author} {\bibfnamefont {M.}~\bibnamefont {Zhu}}, \bibinfo
  {author} {\bibfnamefont {J.}~\bibnamefont {Ayres}}, \bibinfo {author}
  {\bibfnamefont {K.}~\bibnamefont {Kummer}}, \bibinfo {author} {\bibfnamefont
  {F.}~\bibnamefont {Yakhou-Harris}}, \bibinfo {author} {\bibfnamefont {J.~R.}\
  \bibnamefont {Cooper}}, \bibinfo {author} {\bibfnamefont {A.}~\bibnamefont
  {Carrington}},\ and\ \bibinfo {author} {\bibfnamefont {S.~M.}\ \bibnamefont
  {Hayden}},\ }\bibfield  {title} {\bibinfo {title} {Charge density waves and
  fermi surface reconstruction in the clean overdoped cuprate superconductor
  \uppercase{T}l$_2$\uppercase{B}a$_2$\uppercase{C}u\uppercase{O}$_{6+\delta}$},\
  }\href {https://doi.org/10.1038/s41467-022-28124-y} {\bibfield  {journal}
  {\bibinfo  {journal} {Nature Communications}\ }\textbf {\bibinfo {volume}
  {13}},\ \bibinfo {pages} {570} (\bibinfo {year} {2022})}\BibitemShut
  {NoStop}%
\bibitem [{\citenamefont {Chen}\ \emph {et~al.}(2024)\citenamefont {Chen},
  \citenamefont {Gaudet}, \citenamefont {Marcus}, \citenamefont {Nomoto},
  \citenamefont {Chen}, \citenamefont {Tomita}, \citenamefont {Ikhlas},
  \citenamefont {Suzuki}, \citenamefont {Zhao}, \citenamefont {Chen},
  \citenamefont {Strempfer}, \citenamefont {Arita}, \citenamefont {Nakatsuji},\
  and\ \citenamefont {Broholm}}]{PhysRevResearch.6.L032016}%
  \BibitemOpen
  \bibfield  {author} {\bibinfo {author} {\bibfnamefont {Y.}~\bibnamefont
  {Chen}}, \bibinfo {author} {\bibfnamefont {J.}~\bibnamefont {Gaudet}},
  \bibinfo {author} {\bibfnamefont {G.~G.}\ \bibnamefont {Marcus}}, \bibinfo
  {author} {\bibfnamefont {T.}~\bibnamefont {Nomoto}}, \bibinfo {author}
  {\bibfnamefont {T.}~\bibnamefont {Chen}}, \bibinfo {author} {\bibfnamefont
  {T.}~\bibnamefont {Tomita}}, \bibinfo {author} {\bibfnamefont
  {M.}~\bibnamefont {Ikhlas}}, \bibinfo {author} {\bibfnamefont {H.~S.}\
  \bibnamefont {Suzuki}}, \bibinfo {author} {\bibfnamefont {Y.}~\bibnamefont
  {Zhao}}, \bibinfo {author} {\bibfnamefont {W.~C.}\ \bibnamefont {Chen}},
  \bibinfo {author} {\bibfnamefont {J.}~\bibnamefont {Strempfer}}, \bibinfo
  {author} {\bibfnamefont {R.}~\bibnamefont {Arita}}, \bibinfo {author}
  {\bibfnamefont {S.}~\bibnamefont {Nakatsuji}},\ and\ \bibinfo {author}
  {\bibfnamefont {C.}~\bibnamefont {Broholm}},\ }\bibfield  {title} {\bibinfo
  {title} {Intertwined charge and spin density waves in a topological kagome
  material},\ }\href {https://doi.org/10.1103/PhysRevResearch.6.L032016}
  {\bibfield  {journal} {\bibinfo  {journal} {Phys. Rev. Res.}\ }\textbf
  {\bibinfo {volume} {6}},\ \bibinfo {pages} {L032016} (\bibinfo {year}
  {2024})}\BibitemShut {NoStop}%
\bibitem [{\citenamefont {Khasanov}\ \emph {et~al.}(2016)\citenamefont
  {Khasanov}, \citenamefont {Guguchia}, \citenamefont {Maisuradze},
  \citenamefont {Andreica}, \citenamefont {Elender}, \citenamefont {Raselli},
  \citenamefont {Shermadini}, \citenamefont {Goko}, \citenamefont {Knecht},
  \citenamefont {Morenzoni},\ and\ \citenamefont
  {Amato}}]{doi:10.1080/08957959.2016.1173690}%
  \BibitemOpen
  \bibfield  {author} {\bibinfo {author} {\bibfnamefont {R.}~\bibnamefont
  {Khasanov}}, \bibinfo {author} {\bibfnamefont {Z.}~\bibnamefont {Guguchia}},
  \bibinfo {author} {\bibfnamefont {A.}~\bibnamefont {Maisuradze}}, \bibinfo
  {author} {\bibfnamefont {D.}~\bibnamefont {Andreica}}, \bibinfo {author}
  {\bibfnamefont {M.}~\bibnamefont {Elender}}, \bibinfo {author} {\bibfnamefont
  {A.}~\bibnamefont {Raselli}}, \bibinfo {author} {\bibfnamefont
  {Z.}~\bibnamefont {Shermadini}}, \bibinfo {author} {\bibfnamefont
  {T.}~\bibnamefont {Goko}}, \bibinfo {author} {\bibfnamefont {F.}~\bibnamefont
  {Knecht}}, \bibinfo {author} {\bibfnamefont {E.}~\bibnamefont {Morenzoni}},\
  and\ \bibinfo {author} {\bibfnamefont {A.}~\bibnamefont {Amato}},\ }\bibfield
   {title} {\bibinfo {title} {High pressure research using muons at the paul
  scherrer institute},\ }\href {https://doi.org/10.1080/08957959.2016.1173690}
  {\bibfield  {journal} {\bibinfo  {journal} {High Pressure Research}\ }\textbf
  {\bibinfo {volume} {36}},\ \bibinfo {pages} {140} (\bibinfo {year}
  {2016})}\BibitemShut {NoStop}%
\bibitem [{\citenamefont {Khasanov}(2022)}]{10.1063/5.0119840}%
  \BibitemOpen
  \bibfield  {author} {\bibinfo {author} {\bibfnamefont {R.}~\bibnamefont
  {Khasanov}},\ }\bibfield  {title} {\bibinfo {title} {{Perspective on
  muon-spin rotation/relaxation under hydrostatic pressure}},\ }\href
  {https://doi.org/10.1063/5.0119840} {\bibfield  {journal} {\bibinfo
  {journal} {Journal of Applied Physics}\ }\textbf {\bibinfo {volume} {132}}
  (\bibinfo {year} {2022})}\BibitemShut {NoStop}%
\bibitem [{\citenamefont {Z.~Shermadini}\ and\ \citenamefont
  {Amato}(2017)}]{doi:10.1080/08957959.2017.1373773}%
  \BibitemOpen
  \bibfield  {author} {\bibinfo {author} {\bibfnamefont {M.~E. G. S. Z. G.
  K.~K.}\ \bibnamefont {Z.~Shermadini}, \bibfnamefont {R.~Khasanov}}\ and\
  \bibinfo {author} {\bibfnamefont {A.}~\bibnamefont {Amato}},\ }\bibfield
  {title} {\bibinfo {title} {A low-background piston–cylinder-type hybrid
  high pressure cell for muon-spin rotation/relaxation experiments},\ }\href
  {https://doi.org/10.1080/08957959.2017.1373773} {\bibfield  {journal}
  {\bibinfo  {journal} {High Pressure Research}\ }\textbf {\bibinfo {volume}
  {37}},\ \bibinfo {pages} {449} (\bibinfo {year} {2017})}\BibitemShut
  {NoStop}%
\bibitem [{\citenamefont {Suter}\ and\ \citenamefont
  {Wojek}(2012)}]{SUTER201269}%
  \BibitemOpen
  \bibfield  {author} {\bibinfo {author} {\bibfnamefont {A.}~\bibnamefont
  {Suter}}\ and\ \bibinfo {author} {\bibfnamefont {B.}~\bibnamefont {Wojek}},\
  }\bibfield  {title} {\bibinfo {title} {Musrfit: A free platform-independent
  framework for µsr data analysis},\ }\href
  {https://doi.org/https://doi.org/10.1016/j.phpro.2012.04.042} {\bibfield
  {journal} {\bibinfo  {journal} {Physics Procedia}\ }\textbf {\bibinfo
  {volume} {30}},\ \bibinfo {pages} {69} (\bibinfo {year} {2012})},\ \bibinfo
  {note} {12th International Conference on Muon Spin Rotation, Relaxation and
  Resonance ($\mu$SR2011)}\BibitemShut {NoStop}%
\bibitem [{\citenamefont {Paranjpe}\ and\ \citenamefont
  {Dande}(1989)}]{Paranjpe1989}%
  \BibitemOpen
  \bibfield  {author} {\bibinfo {author} {\bibfnamefont {S.~K.}\ \bibnamefont
  {Paranjpe}}\ and\ \bibinfo {author} {\bibfnamefont {Y.~D.}\ \bibnamefont
  {Dande}},\ }\bibfield  {title} {\bibinfo {title} {A neutron diffractometer
  with a linear position sensitive detector},\ }\href
  {https://doi.org/10.1007/BF02845999} {\bibfield  {journal} {\bibinfo
  {journal} {Pramana}\ }\textbf {\bibinfo {volume} {32}},\ \bibinfo {pages}
  {793} (\bibinfo {year} {1989})}\BibitemShut {NoStop}%
\bibitem [{\citenamefont {Paranjpe}\ and\ \citenamefont
  {Yusuf}(2002)}]{doi:10.1080/10448630208218701}%
  \BibitemOpen
  \bibfield  {author} {\bibinfo {author} {\bibfnamefont {S.~K.}\ \bibnamefont
  {Paranjpe}}\ and\ \bibinfo {author} {\bibfnamefont {S.~M.}\ \bibnamefont
  {Yusuf}},\ }\bibfield  {title} {\bibinfo {title} {Neutron powder diffraction
  studies at barc},\ }\href {https://doi.org/10.1080/10448630208218701}
  {\bibfield  {journal} {\bibinfo  {journal} {Neutron News}\ }\textbf {\bibinfo
  {volume} {13}},\ \bibinfo {pages} {36} (\bibinfo {year} {2002})}\BibitemShut
  {NoStop}%
\bibitem [{\citenamefont {Rodriguez-Carvajal}(1990)}]{Rodriguez1990}%
  \BibitemOpen
  \bibfield  {author} {\bibinfo {author} {\bibfnamefont {J.}~\bibnamefont
  {Rodriguez-Carvajal}},\ }\bibfield  {title} {\bibinfo {title} {Fullprof: a
  program for rietveld refinement and pattern matching analysis},\ }in\ \href
  {https://www.ill.eu/sites/fullprof/php/reference.html} {\emph {\bibinfo
  {booktitle} {satellite meeting on powder diffraction of the XV congress of
  the IUCr}}},\ Vol.\ \bibinfo {volume} {127}\ (\bibinfo {organization}
  {Toulouse, France]},\ \bibinfo {year} {1990})\BibitemShut {NoStop}%
\bibitem [{\citenamefont {Kurosawa}\ \emph {et~al.}(2022)\citenamefont
  {Kurosawa}, \citenamefont {Tomita}, \citenamefont {Muhammad}, \citenamefont
  {Fu}, \citenamefont {Sakai},\ and\ \citenamefont
  {Nakatsuji}}]{kurosawa2022chiralanomalydriven}%
  \BibitemOpen
  \bibfield  {author} {\bibinfo {author} {\bibfnamefont {S.}~\bibnamefont
  {Kurosawa}}, \bibinfo {author} {\bibfnamefont {T.}~\bibnamefont {Tomita}},
  \bibinfo {author} {\bibfnamefont {I.}~\bibnamefont {Muhammad}}, \bibinfo
  {author} {\bibfnamefont {M.}~\bibnamefont {Fu}}, \bibinfo {author}
  {\bibfnamefont {A.}~\bibnamefont {Sakai}},\ and\ \bibinfo {author}
  {\bibfnamefont {S.}~\bibnamefont {Nakatsuji}},\ }\bibfield  {title} {\bibinfo
  {title} {Chiral-anomaly-driven magnetotransport in the correlated weyl magnet
  \uppercase{M}n$_3$\uppercase{S}n},\ }\href@noop {} {\  (\bibinfo {year}
  {2022})},\ \Eprint {https://arxiv.org/abs/2204.00882} {arXiv:2204.00882
  [cond-mat.str-el]} \BibitemShut {NoStop}%
\bibitem [{\citenamefont {Taylor}\ \emph {et~al.}(2020)\citenamefont {Taylor},
  \citenamefont {Markou}, \citenamefont {Lesne}, \citenamefont {Sivakumar},
  \citenamefont {Luo}, \citenamefont {Radu}, \citenamefont {Werner},
  \citenamefont {Felser},\ and\ \citenamefont {Parkin}}]{PhysRevB.101.094404}%
  \BibitemOpen
  \bibfield  {author} {\bibinfo {author} {\bibfnamefont {J.~M.}\ \bibnamefont
  {Taylor}}, \bibinfo {author} {\bibfnamefont {A.}~\bibnamefont {Markou}},
  \bibinfo {author} {\bibfnamefont {E.}~\bibnamefont {Lesne}}, \bibinfo
  {author} {\bibfnamefont {P.~K.}\ \bibnamefont {Sivakumar}}, \bibinfo {author}
  {\bibfnamefont {C.}~\bibnamefont {Luo}}, \bibinfo {author} {\bibfnamefont
  {F.}~\bibnamefont {Radu}}, \bibinfo {author} {\bibfnamefont {P.}~\bibnamefont
  {Werner}}, \bibinfo {author} {\bibfnamefont {C.}~\bibnamefont {Felser}},\
  and\ \bibinfo {author} {\bibfnamefont {S.~S.~P.}\ \bibnamefont {Parkin}},\
  }\bibfield  {title} {\bibinfo {title} {Anomalous and topological hall effects
  in epitaxial thin films of the noncollinear antiferromagnet
  ${\mathrm{\uppercase{m}n}}_{3}\mathrm{\uppercase{s}n}$},\ }\href
  {https://doi.org/10.1103/PhysRevB.101.094404} {\bibfield  {journal} {\bibinfo
   {journal} {Phys. Rev. B}\ }\textbf {\bibinfo {volume} {101}},\ \bibinfo
  {pages} {094404} (\bibinfo {year} {2020})}\BibitemShut {NoStop}%
\bibitem [{\citenamefont {Feng}\ \emph {et~al.}(2006)\citenamefont {Feng},
  \citenamefont {Li}, \citenamefont {Ren}, \citenamefont {Li}, \citenamefont
  {Li}, \citenamefont {Li}, \citenamefont {Zhang},\ and\ \citenamefont
  {Zhang}}]{PhysRevB.73.205105}%
  \BibitemOpen
  \bibfield  {author} {\bibinfo {author} {\bibfnamefont {W.~J.}\ \bibnamefont
  {Feng}}, \bibinfo {author} {\bibfnamefont {D.}~\bibnamefont {Li}}, \bibinfo
  {author} {\bibfnamefont {W.~J.}\ \bibnamefont {Ren}}, \bibinfo {author}
  {\bibfnamefont {Y.~B.}\ \bibnamefont {Li}}, \bibinfo {author} {\bibfnamefont
  {W.~F.}\ \bibnamefont {Li}}, \bibinfo {author} {\bibfnamefont
  {J.}~\bibnamefont {Li}}, \bibinfo {author} {\bibfnamefont {Y.~Q.}\
  \bibnamefont {Zhang}},\ and\ \bibinfo {author} {\bibfnamefont {Z.~D.}\
  \bibnamefont {Zhang}},\ }\bibfield  {title} {\bibinfo {title} {Glassy
  ferromagnetism in ${\mathrm{\uppercase{n}i}}_{3}\mathrm{\uppercase{s}n}$-type
  ${\mathrm{\uppercase{m}n}}_{3.1}{\mathrm{\uppercase{s}n}}_{0.9}$},\ }\href
  {https://doi.org/10.1103/PhysRevB.73.205105} {\bibfield  {journal} {\bibinfo
  {journal} {Phys. Rev. B}\ }\textbf {\bibinfo {volume} {73}},\ \bibinfo
  {pages} {205105} (\bibinfo {year} {2006})}\BibitemShut {NoStop}%
\bibitem [{\citenamefont {Singh}\ \emph {et~al.}(2022)\citenamefont {Singh},
  \citenamefont {Jamaluddin}, \citenamefont {Nandy}, \citenamefont {Tokunaga},
  \citenamefont {Avdeev},\ and\ \citenamefont {Nayak}}]{singh2022higher}%
  \BibitemOpen
  \bibfield  {author} {\bibinfo {author} {\bibfnamefont {C.}~\bibnamefont
  {Singh}}, \bibinfo {author} {\bibfnamefont {S.}~\bibnamefont {Jamaluddin}},
  \bibinfo {author} {\bibfnamefont {A.~K.}\ \bibnamefont {Nandy}}, \bibinfo
  {author} {\bibfnamefont {M.}~\bibnamefont {Tokunaga}}, \bibinfo {author}
  {\bibfnamefont {M.}~\bibnamefont {Avdeev}},\ and\ \bibinfo {author}
  {\bibfnamefont {A.~K.}\ \bibnamefont {Nayak}},\ }\href@noop {} {\bibinfo
  {title} {Higher order exchange driven noncoplanar magnetic state and large
  anomalous hall effects in electron doped kagome magnet mn$_3$sn}} (\bibinfo
  {year} {2022}),\ \Eprint {https://arxiv.org/abs/2211.12722} {arXiv:2211.12722
  [cond-mat.mtrl-sci]} \BibitemShut {NoStop}%
\bibitem [{\citenamefont {Getzlaff}(2007)}]{Getzlaff2007-iu}%
  \BibitemOpen
  \bibfield  {author} {\bibinfo {author} {\bibfnamefont {M.}~\bibnamefont
  {Getzlaff}},\ }\href@noop {} {\emph {\bibinfo {title} {Fundamentals of
  magnetism}}},\ \bibinfo {edition} {2008th}\ ed.\ (\bibinfo  {publisher}
  {Springer},\ \bibinfo {address} {Berlin, Germany},\ \bibinfo {year}
  {2007})\BibitemShut {NoStop}%
\bibitem [{\citenamefont {Smidman}\ \emph {et~al.}(2013)\citenamefont
  {Smidman}, \citenamefont {Adroja}, \citenamefont {Hillier}, \citenamefont
  {Chapon}, \citenamefont {Taylor}, \citenamefont {Anand}, \citenamefont
  {Singh}, \citenamefont {Lees}, \citenamefont {Goremychkin}, \citenamefont
  {Koza}, \citenamefont {Krishnamurthy}, \citenamefont {Paul},\ and\
  \citenamefont {Balakrishnan}}]{PhysRevB.88.134416}%
  \BibitemOpen
  \bibfield  {author} {\bibinfo {author} {\bibfnamefont {M.}~\bibnamefont
  {Smidman}}, \bibinfo {author} {\bibfnamefont {D.~T.}\ \bibnamefont {Adroja}},
  \bibinfo {author} {\bibfnamefont {A.~D.}\ \bibnamefont {Hillier}}, \bibinfo
  {author} {\bibfnamefont {L.~C.}\ \bibnamefont {Chapon}}, \bibinfo {author}
  {\bibfnamefont {J.~W.}\ \bibnamefont {Taylor}}, \bibinfo {author}
  {\bibfnamefont {V.~K.}\ \bibnamefont {Anand}}, \bibinfo {author}
  {\bibfnamefont {R.~P.}\ \bibnamefont {Singh}}, \bibinfo {author}
  {\bibfnamefont {M.~R.}\ \bibnamefont {Lees}}, \bibinfo {author}
  {\bibfnamefont {E.~A.}\ \bibnamefont {Goremychkin}}, \bibinfo {author}
  {\bibfnamefont {M.~M.}\ \bibnamefont {Koza}}, \bibinfo {author}
  {\bibfnamefont {V.~V.}\ \bibnamefont {Krishnamurthy}}, \bibinfo {author}
  {\bibfnamefont {D.~M.}\ \bibnamefont {Paul}},\ and\ \bibinfo {author}
  {\bibfnamefont {G.}~\bibnamefont {Balakrishnan}},\ }\bibfield  {title}
  {\bibinfo {title} {Neutron scattering and muon spin relaxation measurements
  of the noncentrosymmetric antiferromagnet $\mathrm{CeCoGe}_3$},\ }\href
  {https://doi.org/10.1103/PhysRevB.88.134416} {\bibfield  {journal} {\bibinfo
  {journal} {Phys. Rev. B}\ }\textbf {\bibinfo {volume} {88}},\ \bibinfo
  {pages} {134416} (\bibinfo {year} {2013})}\BibitemShut {NoStop}%
\bibitem [{\citenamefont {Yaouanc}\ and\ \citenamefont
  {Reotier}(2011)}]{Muon_Book_1}%
  \BibitemOpen
  \bibfield  {author} {\bibinfo {author} {\bibfnamefont {A.}~\bibnamefont
  {Yaouanc}}\ and\ \bibinfo {author} {\bibfnamefont {P.~D.~D.}\ \bibnamefont
  {Reotier}},\ }\href@noop {} {\emph {\bibinfo {title} {Muon Spin Rotation,
  Relaxation, and Resonance: Applications to Condensed Matter}}}\ (\bibinfo
  {publisher} {Oxford University Press},\ \bibinfo {year} {2011})\ pp.\
  \bibinfo {pages} {247--259}\BibitemShut {NoStop}%
\bibitem [{\citenamefont {Andreica}(2001)}]{20.500.11850/145189}%
  \BibitemOpen
  \bibfield  {author} {\bibinfo {author} {\bibfnamefont {D.-A.}\ \bibnamefont
  {Andreica}},\ }\emph {\bibinfo {title} {Magnetic phase diagram in some
  Kondo-lattice compounds. microscopic and macroscopic studies}},\ \href
  {https://doi.org/10.3929/ethz-a-004171645} {\bibinfo {type} {Doctoral
  thesis}},\ \bibinfo  {school} {ETH Zurich}, \bibinfo {address} {Zürich}
  (\bibinfo {year} {2001}),\ \bibinfo {note} {diss., Naturwissenschaften ETH
  Zürich, Nr. 14170, 2001.}\BibitemShut {Stop}%
\bibitem [{\citenamefont {Eyring}\ \emph {et~al.}(2001)\citenamefont {Eyring},
  \citenamefont {Gschneidner},\ and\ \citenamefont {Lander}}]{Eyring2001-lg}%
  \BibitemOpen
  \bibinfo {editor} {\bibfnamefont {L.}~\bibnamefont {Eyring}}, \bibinfo
  {editor} {\bibfnamefont {K.~A.}\ \bibnamefont {Gschneidner}},\ and\ \bibinfo
  {editor} {\bibfnamefont {G.~H.}\ \bibnamefont {Lander}},\ eds.,\ \href@noop
  {} {\emph {\bibinfo {title} {Handbook on the physics and chemistry of rare
  earths: Volume 32}}},\ Handbook on the Physics \& Chemistry of Rare Earths\
  (\bibinfo  {publisher} {North-Holland},\ \bibinfo {address} {Oxford,
  England},\ \bibinfo {year} {2001})\BibitemShut {NoStop}%
\bibitem [{\citenamefont {Schenck}\ \emph {et~al.}(2001)\citenamefont
  {Schenck}, \citenamefont {Andreica}, \citenamefont {Gygax},\ and\
  \citenamefont {Ott}}]{PhysRevB.65.024444}%
  \BibitemOpen
  \bibfield  {author} {\bibinfo {author} {\bibfnamefont {A.}~\bibnamefont
  {Schenck}}, \bibinfo {author} {\bibfnamefont {D.}~\bibnamefont {Andreica}},
  \bibinfo {author} {\bibfnamefont {F.~N.}\ \bibnamefont {Gygax}},\ and\
  \bibinfo {author} {\bibfnamefont {H.~R.}\ \bibnamefont {Ott}},\ }\bibfield
  {title} {\bibinfo {title} {Extreme quantum behavior of positive muons in
  ${\mathrm{\uppercase{c}e\uppercase{a}l}}_{2}$ below 1 k},\ }\href
  {https://doi.org/10.1103/PhysRevB.65.024444} {\bibfield  {journal} {\bibinfo
  {journal} {Phys. Rev. B}\ }\textbf {\bibinfo {volume} {65}},\ \bibinfo
  {pages} {024444} (\bibinfo {year} {2001})}\BibitemShut {NoStop}%
\bibitem [{\citenamefont {Sugiyama}\ \emph {et~al.}(2011)\citenamefont
  {Sugiyama}, \citenamefont {M\aa{}nsson}, \citenamefont {Ofer}, \citenamefont
  {Kamazawa}, \citenamefont {Harada}, \citenamefont {Andreica}, \citenamefont
  {Amato}, \citenamefont {Brewer}, \citenamefont {Ansaldo}, \citenamefont
  {Ohta}, \citenamefont {Michioka},\ and\ \citenamefont
  {Yoshimura}}]{PhysRevB.84.184421}%
  \BibitemOpen
  \bibfield  {author} {\bibinfo {author} {\bibfnamefont {J.}~\bibnamefont
  {Sugiyama}}, \bibinfo {author} {\bibfnamefont {M.}~\bibnamefont
  {M\aa{}nsson}}, \bibinfo {author} {\bibfnamefont {O.}~\bibnamefont {Ofer}},
  \bibinfo {author} {\bibfnamefont {K.}~\bibnamefont {Kamazawa}}, \bibinfo
  {author} {\bibfnamefont {M.}~\bibnamefont {Harada}}, \bibinfo {author}
  {\bibfnamefont {D.}~\bibnamefont {Andreica}}, \bibinfo {author}
  {\bibfnamefont {A.}~\bibnamefont {Amato}}, \bibinfo {author} {\bibfnamefont
  {J.~H.}\ \bibnamefont {Brewer}}, \bibinfo {author} {\bibfnamefont {E.~J.}\
  \bibnamefont {Ansaldo}}, \bibinfo {author} {\bibfnamefont {H.}~\bibnamefont
  {Ohta}}, \bibinfo {author} {\bibfnamefont {C.}~\bibnamefont {Michioka}},\
  and\ \bibinfo {author} {\bibfnamefont {K.}~\bibnamefont {Yoshimura}},\
  }\bibfield  {title} {\bibinfo {title} {Successive magnetic transitions and
  static magnetic order in
  $\uppercase{R}$\uppercase{C}o\uppercase{A}s\uppercase{O}
  ($\uppercase{R}=\text{\uppercase{l}a}$, \uppercase{C}e, \uppercase{P}r,
  \uppercase{N}d, \uppercase{S}m, \uppercase{G}d) confirmed by muon-spin
  rotation and relaxation},\ }\href
  {https://doi.org/10.1103/PhysRevB.84.184421} {\bibfield  {journal} {\bibinfo
  {journal} {Phys. Rev. B}\ }\textbf {\bibinfo {volume} {84}},\ \bibinfo
  {pages} {184421} (\bibinfo {year} {2011})}\BibitemShut {NoStop}%
\bibitem [{\citenamefont {Zhu}\ \emph {et~al.}(2022)\citenamefont {Zhu},
  \citenamefont {Zhang}, \citenamefont {Gawryluk}, \citenamefont {Zhen},
  \citenamefont {Yu}, \citenamefont {Ju}, \citenamefont {Xie}, \citenamefont
  {Jiang}, \citenamefont {Cheng}, \citenamefont {Xu}, \citenamefont {Shi},
  \citenamefont {Pomjakushina}, \citenamefont {Zhan}, \citenamefont {Shiroka},\
  and\ \citenamefont {Shang}}]{PhysRevB.105.014423}%
  \BibitemOpen
  \bibfield  {author} {\bibinfo {author} {\bibfnamefont {X.~Y.}\ \bibnamefont
  {Zhu}}, \bibinfo {author} {\bibfnamefont {H.}~\bibnamefont {Zhang}}, \bibinfo
  {author} {\bibfnamefont {D.~J.}\ \bibnamefont {Gawryluk}}, \bibinfo {author}
  {\bibfnamefont {Z.~X.}\ \bibnamefont {Zhen}}, \bibinfo {author}
  {\bibfnamefont {B.~C.}\ \bibnamefont {Yu}}, \bibinfo {author} {\bibfnamefont
  {S.~L.}\ \bibnamefont {Ju}}, \bibinfo {author} {\bibfnamefont
  {W.}~\bibnamefont {Xie}}, \bibinfo {author} {\bibfnamefont {D.~M.}\
  \bibnamefont {Jiang}}, \bibinfo {author} {\bibfnamefont {W.~J.}\ \bibnamefont
  {Cheng}}, \bibinfo {author} {\bibfnamefont {Y.}~\bibnamefont {Xu}}, \bibinfo
  {author} {\bibfnamefont {M.}~\bibnamefont {Shi}}, \bibinfo {author}
  {\bibfnamefont {E.}~\bibnamefont {Pomjakushina}}, \bibinfo {author}
  {\bibfnamefont {Q.~F.}\ \bibnamefont {Zhan}}, \bibinfo {author}
  {\bibfnamefont {T.}~\bibnamefont {Shiroka}},\ and\ \bibinfo {author}
  {\bibfnamefont {T.}~\bibnamefont {Shang}},\ }\bibfield  {title} {\bibinfo
  {title} {Spin order and fluctuations in the $\mathrm{EuAl}_4$ and
  $\mathrm{EuGa}_4$ topological antiferromagnets: A
  $\ensuremath{\mu}\mathrm{SR}$ study},\ }\href
  {https://doi.org/10.1103/PhysRevB.105.014423} {\bibfield  {journal} {\bibinfo
   {journal} {Phys. Rev. B}\ }\textbf {\bibinfo {volume} {105}},\ \bibinfo
  {pages} {014423} (\bibinfo {year} {2022})}\BibitemShut {NoStop}%
\bibitem [{\citenamefont {Zorko}\ \emph {et~al.}(2019)\citenamefont {Zorko},
  \citenamefont {Pregelj}, \citenamefont {Klanj\ifmmode~\check{s}\else
  \v{s}\fi{}ek}, \citenamefont {Gomil\ifmmode~\check{s}\else \v{s}\fi{}ek},
  \citenamefont {Jagli\ifmmode \check{c}\else
  \v{c}\fi{}i\ifmmode~\acute{c}\else \'{c}\fi{}}, \citenamefont {Lord},
  \citenamefont {Verezhak}, \citenamefont {Shang}, \citenamefont {Sun},\ and\
  \citenamefont {Mi}}]{PhysRevB.99.214441}%
  \BibitemOpen
  \bibfield  {author} {\bibinfo {author} {\bibfnamefont {A.}~\bibnamefont
  {Zorko}}, \bibinfo {author} {\bibfnamefont {M.}~\bibnamefont {Pregelj}},
  \bibinfo {author} {\bibfnamefont {M.}~\bibnamefont
  {Klanj\ifmmode~\check{s}\else \v{s}\fi{}ek}}, \bibinfo {author}
  {\bibfnamefont {M.}~\bibnamefont {Gomil\ifmmode~\check{s}\else
  \v{s}\fi{}ek}}, \bibinfo {author} {\bibfnamefont {Z.}~\bibnamefont
  {Jagli\ifmmode \check{c}\else \v{c}\fi{}i\ifmmode~\acute{c}\else
  \'{c}\fi{}}}, \bibinfo {author} {\bibfnamefont {J.~S.}\ \bibnamefont {Lord}},
  \bibinfo {author} {\bibfnamefont {J.~A.~T.}\ \bibnamefont {Verezhak}},
  \bibinfo {author} {\bibfnamefont {T.}~\bibnamefont {Shang}}, \bibinfo
  {author} {\bibfnamefont {W.}~\bibnamefont {Sun}},\ and\ \bibinfo {author}
  {\bibfnamefont {J.-X.}\ \bibnamefont {Mi}},\ }\bibfield  {title} {\bibinfo
  {title} {Coexistence of magnetic order and persistent spin dynamics in a
  quantum kagome antiferromagnet with no intersite mixing},\ }\href
  {https://doi.org/10.1103/PhysRevB.99.214441} {\bibfield  {journal} {\bibinfo
  {journal} {Phys. Rev. B}\ }\textbf {\bibinfo {volume} {99}},\ \bibinfo
  {pages} {214441} (\bibinfo {year} {2019})}\BibitemShut {NoStop}%
\bibitem [{\citenamefont {Kim}\ and\ \citenamefont
  {Liu}(2023)}]{PhysRevB.107.205130}%
  \BibitemOpen
  \bibfield  {author} {\bibinfo {author} {\bibfnamefont {D.}~\bibnamefont
  {Kim}}\ and\ \bibinfo {author} {\bibfnamefont {F.}~\bibnamefont {Liu}},\
  }\bibfield  {title} {\bibinfo {title} {Realization of flat bands by lattice
  intercalation in kagome metals},\ }\href
  {https://doi.org/10.1103/PhysRevB.107.205130} {\bibfield  {journal} {\bibinfo
   {journal} {Phys. Rev. B}\ }\textbf {\bibinfo {volume} {107}},\ \bibinfo
  {pages} {205130} (\bibinfo {year} {2023})}\BibitemShut {NoStop}%
\bibitem [{\citenamefont {Okamoto}\ \emph {et~al.}(2022)\citenamefont
  {Okamoto}, \citenamefont {Mohanta}, \citenamefont {Dagotto},\ and\
  \citenamefont {Sheng}}]{Okamoto2022}%
  \BibitemOpen
  \bibfield  {author} {\bibinfo {author} {\bibfnamefont {S.}~\bibnamefont
  {Okamoto}}, \bibinfo {author} {\bibfnamefont {N.}~\bibnamefont {Mohanta}},
  \bibinfo {author} {\bibfnamefont {E.}~\bibnamefont {Dagotto}},\ and\ \bibinfo
  {author} {\bibfnamefont {D.~N.}\ \bibnamefont {Sheng}},\ }\bibfield  {title}
  {\bibinfo {title} {Topological flat bands in a kagome lattice multiorbital
  system},\ }\href {https://doi.org/10.1038/s42005-022-00969-1} {\bibfield
  {journal} {\bibinfo  {journal} {Communications Physics}\ }\textbf {\bibinfo
  {volume} {5}},\ \bibinfo {pages} {198} (\bibinfo {year} {2022})}\BibitemShut
  {NoStop}%
\bibitem [{\citenamefont {Zhang}\ \emph {et~al.}(2013)\citenamefont {Zhang},
  \citenamefont {Yan}, \citenamefont {Wu}, \citenamefont {Kübler},
  \citenamefont {Kreiner}, \citenamefont {Parkin},\ and\ \citenamefont
  {Felser}}]{Zhang_2013}%
  \BibitemOpen
  \bibfield  {author} {\bibinfo {author} {\bibfnamefont {D.}~\bibnamefont
  {Zhang}}, \bibinfo {author} {\bibfnamefont {B.}~\bibnamefont {Yan}}, \bibinfo
  {author} {\bibfnamefont {S.-C.}\ \bibnamefont {Wu}}, \bibinfo {author}
  {\bibfnamefont {J.}~\bibnamefont {Kübler}}, \bibinfo {author} {\bibfnamefont
  {G.}~\bibnamefont {Kreiner}}, \bibinfo {author} {\bibfnamefont {S.~S.~P.}\
  \bibnamefont {Parkin}},\ and\ \bibinfo {author} {\bibfnamefont
  {C.}~\bibnamefont {Felser}},\ }\bibfield  {title} {\bibinfo {title}
  {First-principles study of the structural stability of cubic, tetragonal and
  hexagonal phases in $\mathrm{Mn}_3\mathrm{Z} (\mathrm{Z}=\mathrm{Ga},\mathrm{
  Sn}$ and $\mathrm{Ge}$) heusler compounds},\ }\href
  {https://doi.org/10.1088/0953-8984/25/20/206006} {\bibfield  {journal}
  {\bibinfo  {journal} {Journal of Physics: Condensed Matter}\ }\textbf
  {\bibinfo {volume} {25}},\ \bibinfo {pages} {206006} (\bibinfo {year}
  {2013})}\BibitemShut {NoStop}%
\bibitem [{\citenamefont {Kren}\ \emph
  {et~al.}(1975{\natexlab{b}})\citenamefont {Kren}, \citenamefont {Paitz},
  \citenamefont {Zimmer},\ and\ \citenamefont {Zsoldos}}]{KREN1975226}%
  \BibitemOpen
  \bibfield  {author} {\bibinfo {author} {\bibfnamefont {E.}~\bibnamefont
  {Kren}}, \bibinfo {author} {\bibfnamefont {J.}~\bibnamefont {Paitz}},
  \bibinfo {author} {\bibfnamefont {G.}~\bibnamefont {Zimmer}},\ and\ \bibinfo
  {author} {\bibfnamefont {E.}~\bibnamefont {Zsoldos}},\ }\bibfield  {title}
  {\bibinfo {title} {Study of the magnetic phase transformation in the
  $\mathrm{Mn}_3\mathrm{Sn}$ phase},\ }\href
  {https://doi.org/https://doi.org/10.1016/0378-4363(75)90066-2} {\bibfield
  {journal} {\bibinfo  {journal} {Physica B+C}\ }\textbf {\bibinfo {volume}
  {80}},\ \bibinfo {pages} {226} (\bibinfo {year}
  {1975}{\natexlab{b}})}\BibitemShut {NoStop}%
\bibitem [{\citenamefont {Tomiyoshi}\ and\ \citenamefont
  {Yamaguchi}(1982{\natexlab{b}})}]{doi:10.1143/JPSJ.51.2478}%
  \BibitemOpen
  \bibfield  {author} {\bibinfo {author} {\bibfnamefont {S.}~\bibnamefont
  {Tomiyoshi}}\ and\ \bibinfo {author} {\bibfnamefont {Y.}~\bibnamefont
  {Yamaguchi}},\ }\bibfield  {title} {\bibinfo {title} {Magnetic structure and
  weak ferromagnetism of mn3sn studied by polarized neutron diffraction},\
  }\href {https://doi.org/10.1143/JPSJ.51.2478} {\bibfield  {journal} {\bibinfo
   {journal} {Journal of the Physical Society of Japan}\ }\textbf {\bibinfo
  {volume} {51}},\ \bibinfo {pages} {2478} (\bibinfo {year}
  {1982}{\natexlab{b}})}\BibitemShut {NoStop}%
\bibitem [{\citenamefont {Kresse}\ and\ \citenamefont
  {Furthm\"uller}(1996)}]{PhysRevB.54.11169}%
  \BibitemOpen
  \bibfield  {author} {\bibinfo {author} {\bibfnamefont {G.}~\bibnamefont
  {Kresse}}\ and\ \bibinfo {author} {\bibfnamefont {J.}~\bibnamefont
  {Furthm\"uller}},\ }\bibfield  {title} {\bibinfo {title} {Efficient iterative
  schemes for ab initio total-energy calculations using a plane-wave basis
  set},\ }\href {https://doi.org/10.1103/PhysRevB.54.11169} {\bibfield
  {journal} {\bibinfo  {journal} {Phys. Rev. B}\ }\textbf {\bibinfo {volume}
  {54}},\ \bibinfo {pages} {11169} (\bibinfo {year} {1996})}\BibitemShut
  {NoStop}%
\bibitem [{\citenamefont {Kresse}\ and\ \citenamefont
  {Joubert}(1999)}]{PhysRevB.59.1758}%
  \BibitemOpen
  \bibfield  {author} {\bibinfo {author} {\bibfnamefont {G.}~\bibnamefont
  {Kresse}}\ and\ \bibinfo {author} {\bibfnamefont {D.}~\bibnamefont
  {Joubert}},\ }\bibfield  {title} {\bibinfo {title} {From ultrasoft
  pseudopotentials to the projector augmented-wave method},\ }\href
  {https://doi.org/10.1103/PhysRevB.59.1758} {\bibfield  {journal} {\bibinfo
  {journal} {Phys. Rev. B}\ }\textbf {\bibinfo {volume} {59}},\ \bibinfo
  {pages} {1758} (\bibinfo {year} {1999})}\BibitemShut {NoStop}%
\bibitem [{\citenamefont {Perdew}\ \emph {et~al.}(1996)\citenamefont {Perdew},
  \citenamefont {Burke},\ and\ \citenamefont
  {Ernzerhof}}]{PhysRevLett.77.3865}%
  \BibitemOpen
  \bibfield  {author} {\bibinfo {author} {\bibfnamefont {J.~P.}\ \bibnamefont
  {Perdew}}, \bibinfo {author} {\bibfnamefont {K.}~\bibnamefont {Burke}},\ and\
  \bibinfo {author} {\bibfnamefont {M.}~\bibnamefont {Ernzerhof}},\ }\bibfield
  {title} {\bibinfo {title} {Generalized gradient approximation made simple},\
  }\href {https://doi.org/10.1103/PhysRevLett.77.3865} {\bibfield  {journal}
  {\bibinfo  {journal} {Phys. Rev. Lett.}\ }\textbf {\bibinfo {volume} {77}},\
  \bibinfo {pages} {3865} (\bibinfo {year} {1996})}\BibitemShut {NoStop}%
\bibitem [{\citenamefont {Hobbs}\ \emph {et~al.}(2000)\citenamefont {Hobbs},
  \citenamefont {Kresse},\ and\ \citenamefont {Hafner}}]{PhysRevB.62.11556}%
  \BibitemOpen
  \bibfield  {author} {\bibinfo {author} {\bibfnamefont {D.}~\bibnamefont
  {Hobbs}}, \bibinfo {author} {\bibfnamefont {G.}~\bibnamefont {Kresse}},\ and\
  \bibinfo {author} {\bibfnamefont {J.}~\bibnamefont {Hafner}},\ }\bibfield
  {title} {\bibinfo {title} {Fully unconstrained noncollinear magnetism within
  the projector augmented-wave method},\ }\href
  {https://doi.org/10.1103/PhysRevB.62.11556} {\bibfield  {journal} {\bibinfo
  {journal} {Phys. Rev. B}\ }\textbf {\bibinfo {volume} {62}},\ \bibinfo
  {pages} {11556} (\bibinfo {year} {2000})}\BibitemShut {NoStop}%
\bibitem [{\citenamefont {Ganose}\ \emph {et~al.}(2021)\citenamefont {Ganose},
  \citenamefont {Searle}, \citenamefont {Jain},\ and\ \citenamefont
  {Griffin}}]{Ganose2021}%
  \BibitemOpen
  \bibfield  {author} {\bibinfo {author} {\bibfnamefont {A.~M.}\ \bibnamefont
  {Ganose}}, \bibinfo {author} {\bibfnamefont {A.}~\bibnamefont {Searle}},
  \bibinfo {author} {\bibfnamefont {A.}~\bibnamefont {Jain}},\ and\ \bibinfo
  {author} {\bibfnamefont {S.~M.}\ \bibnamefont {Griffin}},\ }\bibfield
  {title} {\bibinfo {title} {Ifermi: A python library for fermi surface
  generation and analysis},\ }\href {https://doi.org/10.21105/joss.03089}
  {\bibfield  {journal} {\bibinfo  {journal} {Journal of Open Source Software}\
  }\textbf {\bibinfo {volume} {6}},\ \bibinfo {pages} {3089} (\bibinfo {year}
  {2021})}\BibitemShut {NoStop}%
\bibitem [{\citenamefont {Theuss}\ \emph {et~al.}(2022)\citenamefont {Theuss},
  \citenamefont {Ghosh}, \citenamefont {Chen}, \citenamefont {Tchernyshyov},
  \citenamefont {Nakatsuji},\ and\ \citenamefont
  {Ramshaw}}]{PhysRevB.105.174430}%
  \BibitemOpen
  \bibfield  {author} {\bibinfo {author} {\bibfnamefont {F.}~\bibnamefont
  {Theuss}}, \bibinfo {author} {\bibfnamefont {S.}~\bibnamefont {Ghosh}},
  \bibinfo {author} {\bibfnamefont {T.}~\bibnamefont {Chen}}, \bibinfo {author}
  {\bibfnamefont {O.}~\bibnamefont {Tchernyshyov}}, \bibinfo {author}
  {\bibfnamefont {S.}~\bibnamefont {Nakatsuji}},\ and\ \bibinfo {author}
  {\bibfnamefont {B.~J.}\ \bibnamefont {Ramshaw}},\ }\bibfield  {title}
  {\bibinfo {title} {Strong magnetoelastic coupling in ${\mathrm{mn}}_{3}x$
  ($x=\mathrm{Ge}$, sn)},\ }\href {https://doi.org/10.1103/PhysRevB.105.174430}
  {\bibfield  {journal} {\bibinfo  {journal} {Phys. Rev. B}\ }\textbf {\bibinfo
  {volume} {105}},\ \bibinfo {pages} {174430} (\bibinfo {year}
  {2022})}\BibitemShut {NoStop}%
\bibitem [{\citenamefont {Bonfà}\ \emph {et~al.}()\citenamefont {Bonfà},
  \citenamefont {Onuorah},\ and\ \citenamefont
  {Renzi}}]{doi:10.7566/JPSCP.21.011052}%
  \BibitemOpen
  \bibfield  {author} {\bibinfo {author} {\bibfnamefont {P.}~\bibnamefont
  {Bonfà}}, \bibinfo {author} {\bibfnamefont {I.~J.}\ \bibnamefont
  {Onuorah}},\ and\ \bibinfo {author} {\bibfnamefont {R.~D.}\ \bibnamefont
  {Renzi}},\ }\bibinfo {title} {Introduction and a quick look at muesr, the
  magnetic structure and muon embedding site refinement suite},\ in\ \href
  {https://doi.org/10.7566/JPSCP.21.011052} {\emph {\bibinfo {booktitle}
  {Proceedings of the 14th International Conference on Muon Spin Rotation,
  Relaxation and Resonance ($\mu$SR2017)}}},\ \Eprint
  {https://arxiv.org/abs/https://journals.jps.jp/doi/pdf/10.7566/JPSCP.21.011052}
  {https://journals.jps.jp/doi/pdf/10.7566/JPSCP.21.011052} \BibitemShut
  {NoStop}%
\bibitem [{\citenamefont {Kubo}\ and\ \citenamefont
  {Toyabe}(1967)}]{kubo1967magnetic}%
  \BibitemOpen
  \bibfield  {author} {\bibinfo {author} {\bibfnamefont {R.}~\bibnamefont
  {Kubo}}\ and\ \bibinfo {author} {\bibfnamefont {T.}~\bibnamefont {Toyabe}},\
  }\bibfield  {title} {\bibinfo {title} {Magnetic resonance and relaxation},\
  }in\ \href@noop {} {\emph {\bibinfo {booktitle} {Proceedings of the XIVth
  Colloque Amp{\`e}re}}}\ (\bibinfo {organization} {North-Holland},\ \bibinfo
  {year} {1967})\BibitemShut {NoStop}%
\end{thebibliography}%
\end{document}